\documentclass[thmsa,11pt]{article}%
\usepackage{amsmath}
\usepackage{graphicx}
\usepackage{amsfonts}
\usepackage{amssymb}%
\providecommand{\U}[1]{\protect\rule{.1in}{.1in}}
\begin{document}

\author{J. Doyne Farmer\thanks{Santa Fe Institute, 1399 Hyde Park Rd., Santa Fe NM
87501 and LUISS Guido Carli, Viale Pola 12, 00198, Roma, Italy} \ and John
Geanakoplos\thanks{James Tobin Professor of Economics, Yale University, New
Haven CT, and Santa Fe Institute}}
\title{The virtues and vices of equilibrium\\and the future of financial economics}
\date{\today}
\maketitle

\begin{abstract}
The use of equilibrium models in economics springs from the desire for
parsimonious models of economic phenomena that take human reasoning into
account. This approach has been the cornerstone of modern economic theory. We
explain why this is so, extolling the virtues of equilibrium theory; then we
present a critique and describe why this approach is inherently limited, and
why economics needs to move in new directions if it is to continue to make
progress. We stress that this shouldn't be a question of dogma, but should be
resolved empirically. There are situations where equilibrium models provide
useful predictions and there are situations where they can never provide
useful predictions. There are also many situations where the jury is still
out, i.e., where so far they fail to provide a good description of the world,
but where proper extensions might change this. Our goal is to convince the
skeptics that equilibrium models can be useful, but also to make traditional
economists more aware of the limitations of equilibrium models. We sketch some
alternative approaches and discuss why they should play an important role in
future research in economics.



\end{abstract}
\tableofcontents

\break

\vspace{.5in}

\section{Introduction}

The concept of equilibrium has dominated economics and finance for at least
fifty years. Motivated by the sound desire to find a parsimonious description
of the world, equilibrium theory has had an enormous impact on the way that
economists think, and indeed in many respects defines the way economists
think. Nonetheless, its empirical validity and the extent of its scope still
remains a matter of debate. Its proponents argue that it has been enormously
successful, and that it at least qualitatively explains many aspects of real
economic phenomena. Its detractors argue that when one probes to the bottom,
most of its predictions are essentially unfalsifiable and therefore are not in
fact testable scientific theories. We attempt to give some clarity to this
debate. Our view is that while equilibrium theory is useful, there are
inherent limitations to what it can ever achieve. Economists must expand the
scope of equilibrium theory, explore entirely new approaches, and combine
equilibrium methods with new approaches.

This paper is the outcome of an eight year conversation between an economist
and a physicist. Both of us have been involved in developing trading
strategies for hedge funds, giving us a deep appreciation of the difference
between theories that are empirically useful and those that are merely
aesthetically pleasing. Our hedge funds use completely different strategies.
The strategy the economist developed uses equilibrium methods with behavioral
modifications to trade mortgage-backed securities; the strategy the physicist
developed uses time-series methods based on historical patterns to trade
stocks, an approach that is in some sense the antithesis of the equilibrium
method. Both strategies have been highly successful. We initially came at the
concept of equilibrium from very different points of view, one very supportive
and the other very skeptical. We have had many arguments over the last eight
years. Surprisingly, we have more or less come to agree on the advantages and
disadvantages of the equilibrium approach and the need to go beyond it. The
view presented here is the result of this dialogue.

This paper is organized as follows: In Section~\ref{EqSummary} we present a
short qualitative review of what equilibrium theory is and what it is not, and
in Section~\ref{efficiency} we discuss the related idea of market efficiency
and its importance in finance. In Section~\ref{virtues} we present what we
think are the accomplishments and strengths of equilibrium theory, and in
Section~\ref{limitations} we discuss its inherent limitations from a
theoretical perspective. In Section~\ref{empiricalEvidence} we review the
empirical evidence for and against equilibrium models. Then in
Section~\ref{motivation} we develop some further motivation for
non-equilibrium models, illustrated by a few problems that we think are
inherently out of equilibrium. We also compare equilibrium in physics and
economics and use non-equilibrium models in physics to motivate corresponding
models in economics. In Section~\ref{beyond} we review several alternatives
approaches, including now established approaches such as behavioral and
experimental economics. We also describe some less established approaches to
dealing with bounded rationality, specialization and heterogeneous agents. We
make a distinction between the structural properties of a model (those depend
on the structure of institutions and interactions) vs. the strategic
properties (those that depend on the strategies of agents) and discuss
problems in which structure dominates (so equilibrium may not be very
important). We also discuss how ideas from biology may be useful for
understanding markets. Finally in Section~\ref{conclusions} we present some conclusions.

\section{What is an equilibrium theory?\label{EqSummary}}

In this section we explain the basic idea and the motivation for the
neoclassical concept of economic equilibrium, and discuss some variations on
the basic theme.

Market equilibrium models are designed to describe the market interactions of
rational agents in an economy. The prototype is the general equilibrium theory
of Arrow and Debrieu \cite{Arrow54,Debreu59}. The \textit{economy} consists of
a set of goods, such as seed, corn, apples, and labor time, and a set of
\textit{agents} (households) who decide what to buy and sell, how and what to
produce, and how much to consume of each good. Agents are characterized by
their endowments, technologies, and utilities. The \textit{endowment} of an
agent is the set of all goods she inherits: for example, her ability to labor,
the apples on the trees in her backyard and so on. The \textit{technology} of
an agent is her collection of recipes for transforming goods into others, like
seed into corn. She might buy the seed and hire the labor to do the job, or
use her own seed and labor. Some recipes are better than others, and not all
agents have access to the same technologies. The \textit{utility} of an agent
describes how much happiness she gets out of consuming the goods. The agent
does not get pleasure out of each good separately, but only in the context of
all the goods she consumes taken together.

The goal is to model the decisions of agents incentivized solely by their own
selfish goals, and to deduce the consequences for the production, consumption,
and pricing of each good. The model of Arrow and Debrieu is based on four assumptions:

\begin{enumerate}
\item \textit{Perfect competition} (price taking). Each agent is small enough
to have a negligible effect on prices. Agents \textquotedblleft take" the
prices as given, in the sense that they are offered prices at which they can
buy or sell as much as they want, but the agents cannot do anything to change
the prices.

\item \textit{Agent optimization} of utility. At the given prices each agent
independently chooses what to buy, what and how to produce, and what to sell
so as to maximize her utility. There is no free lunch -- the cost of all
purchases must be financed by sales. An agent can only buy corn, for example,
by selling an equal value of her labor, or apples, or some other good from her endowment.

\item \textit{Market clearing}. The aggregate demand, i.e. the total quantity
that people wish to buy, must be equal to the aggregate supply, i.e. the total
quantity that people wish to sell.

\item \textit{Rational expectations}. Agents make rational decisions based on
perfect information. They have perfect models of the world. They know the
prices of every good before they decide how much to buy or sell of any good.
\end{enumerate}

To actually use this theory one must of course make it concrete by specifying
the set of goods in the economy, how production depends on labor and other
inputs, and a functional form for the utilities. General equilibrium theory
was revolutionary because it provided closure, giving a framework connecting
the different components of the economy to each other and providing a minimum
set of assumptions necessary to get a solution. The behavior of markets is an
extraordinarily complicated and subtle phenomenon, but neoclassical economists
believe that most of it can be explained on the basis of equilibrium.

The advent of general equilibrium theory marked a major transition in the
discipline of economics. It gave hope for a quantitative explanation of the
properties of the economy in one grand theory, causing a sea change in the way
economics is done and in the kind of people that do it. Mainstream economics
shifted from a largely qualitative pursuit, often called \textquotedblleft
political economy", to a highly mathematical field in which papers are often
published in theorem-proof format.

We do not have the space to describe the many applications and consequences of
equilibrium in any detail. The most important consequence for our purposes is
efficiency, which we shall come to shortly. In the remainder of this section
we discuss some logical questions one might raise about whether equilibrium
exists and how and whether it is attainable. Then we give a few examples of
how equilibrium models are being continuously extended to include more
phenomena, and develop the concept of financial equilibrium, our central topic here.

\subsection{Existence of equilibrium and fixed points}

In equilibrium everybody's plans are fulfilled. An agent who plans to buy
three ears of corn by selling two apples at the given prices will find agents
who want to sell him the three ears of corn and others who want to buy his two
apples. A mother who goes to the store to buy her children bread will find as
much as she wants to buy at the going prices on the shelves; yet at the end of
the day the store will not be left with extra bread that goes to waste. One
might well wonder how this could happen without any central coordinator. In
short, why should there be an equilibrium? This is called the existence problem.

In 1954 Arrow, Debreu, and McKenzie simultaneously showed that there always is
an equilibrium, no matter what the endowments and technologies and utilities,
provided that each utility displays diminishing marginal utility of
consumption and each technology displays diminishing marginal product. The
margin is a very famous concept in economics that was discovered by Jevons,
Menger,and Walras in the so-called marginal revolution of 1871. Apples display
diminishing marginal utility if the more apples an agent consumes, the less
additional happiness he achieves from one more apple. Diminishing marginal
product means that the more a good like labor is used in production, the less
extra output comes from one additional unit (hour) of input.

The method of proof used by all these authors was first to guess a vector of
prices, one for each good, then to compute demand and supply at these prices,
and then to increase the prices for the goods with excess demand and to
decrease the prices of goods with excess supply. This defines a map from price
vectors to price vectors. A price vector generates an equilibrium if and only
if it is mapped into itself by this map (for then there is no excess demand or
excess supply). By a clever use of Brouwer's fixed point theorem, it was shown
that there must always be a fixed point and therefore an equilibrium.

\subsection{Getting to Equilibrium}

\subsubsection{Tatonnement}

Who sets the equilibrium prices? How does the economy get to equilibrium? Just
because there is an equilibrium does not mean it is easy to find. One
suggestion of the marginalist Leon Walras was that the equilibrium map could
be iterated over and over. Starting from a guess, one could keep increasing
and decreasing prices depending on whether there was too much demand or
supply. He called this groping (tatonnement in French) toward equilibrium.
Unfortunately, after many years of analysis, it became clear that in general
this tatonnement can cycle without ever getting close to equilibium.

\subsubsection{Omniscience}

An alternative is that the agents know the characteristics of their
competitors and know that they are rational and will not sell their goods at
less than they could get elsewhere; if it is common knowledge that everybody
is rational, and if the characteristics are common knowledge, then each agent
will himself solve for equilibrium and the prices will simply emerge.

This explanation for the emergence of prices has an analogue in Nash
equilibrium in game theory. A game is a contest in which each player makes a
move and receives a payoff that depends on this move as well as the moves of
the other players. The question is how does a player know what move to make
without knowing what moves his opponents will make? And how can he guess what
they will do without knowing what they think he will do? This circular
conundrum was resolved by Nash in 1951 who basically showed that there is
always at least one self-consistent fixed point in the space of strategies. To
be more specific, a \textit{Nash equilibrium} is an assignment of moves to
each player such that if each player thought the others would follow their
assignments, then he would want to follow his. In game theory the question
also arises: who announces this assignment? If the players are all perfectly
logical, and are aware of all their characteristics, they should deduce the assignment.

\subsection{Financial Equilibrium Models}

\subsubsection{Time, Uncertainty, and Securities}

One of the main limitations of general equilibrium theory (as described so
far) is that it is cast in a non-temporal setting. There is only one time
period, so agents don't need to worry about the future. They don't look ahead
and they don't speculate. This simplifies matters enormously but comes at the
heavy cost that the theory is powerless to deal with temporal change. Another
limitation to the theory so far is that agents only trade goods. In the real
world they trade stocks and bonds and other financial securities. Financial
equilibrium extends general equilibrium to allow for time, uncertainty, and
financial securities.

Uncertainty about the future is modeled in terms of \textit{states of nature}
that unfold as the nodes in a tree \cite{Arrow53,Debreu59}. Each state
represents the condition of things like the weather or political events whose
future behavior is unknown, but exogenous to the economy. States are given,
and while they can affect the economy, the economy does not affect them. A
change in the state of nature is often called a \textit{shock}. The agents in
the economy do not know the future states of nature, but they do know the tree
from which the future states of nature are drawn and the probability for
reaching each node. For example, suppose we assume that the dividends of an
agricultural company are made once a year and take on only one of two possible
values, \textquotedblleft high" or \textquotedblleft low", depending on the
weather that year. Then we can organize these as a binary tree in which the
$n^{th}$ level with $2^{n}$ nodes corresponds to the $n^{th}$ year in the
future, and a path from the beginning to the end of the tree corresponds to a
particular sequence of high and low dividends.\footnote{
Financial models can also be formulated with continuous states and/or in
continuous time.}
If the situation was more complicated so that we had to
separate the weather in the western hemisphere from the eastern hemisphere,
then each node would have four branches instead of two and there would be
$4^{n}$ paths. The agents assign probabilities to each branch of the tree,
giving a probability for reaching each node and thus a probability measure on
the terminal states in the tree.

Financial models also extend the notion of equilibrium by generalizing the
notion of what can be traded to include \textit{securities}, which are
promises to deliver certain goods at certain points in time. Such promises can
be contingent on future states of nature. Examples are stocks (which promise
future dividends that change depending on the profits of the firm), bonds
(which promise the same amount each period over a fixed maturity) and so on.
The existence of such securities is important because it allows agents to
decrease their risk. It also allows them to speculate in ways that they might
not otherwise be able to. The implication of such securities for social
welfare has proved to be highly controversial.

In the general equilibrium model each agent is aware of her total consumption
of every good, and makes her decisions in that context. That is why every
decision is a marginal decision: an increase in her consumption of one good
must be evaluated in light of the decrease in her consumption of some other
good that might be required to balance her budget. In the financial
equilibrium extension we are describing, each agent is assumed to be aware of
what her consumption will turn out to be at every node, and to compute her
utility on that basis. One popular way to make that calculation is to evaluate
the utility of consumption at any given node exactly as agents evaluated their
utility of consumption in the general equilibrium model (which is basically a
special case of the financial model with one node). One could then multiply
the utility of consumption at a given node by the probability of ever reaching
the node. Finally, one could discount that number again (i.e. multiply it by a
factor less than 1), depending on how far away from the origin of the tree the
node is, to account for the impatience of consumers. Summing over all the
nodes gives the expected discounted utility of consumption. An agent who
contemplates increasing her consumption of some good in some node in the
financial model must evaluate her utility in light of the decrease in the
holding of some financial security that might then be required for her to
balance her budget, including the consequent loss of consumption in future
states where that security would have paid dividends.


\subsubsection{Financial Equilibrium}

Once the model is extended to allow for the tree of states of nature, and the
trading of securities, we need to extend the notion of equilibrium. This is
done by following principles (1 - 4) above. Prices must now be given at every
node of the tree. When agents optimize they must choose an entire action plan
contemplating what they will buy and sell at each node in the tree; part of
their contemplated trades will involve securities, and the other part
commodities.\footnote{
As the path of future states unfolds the agents will carry out the trades they
anticipated making on that path. Of course only one path will eventually
materialize, and the agents will never get a chance to implement the plans
they had made for unrealized branches.}
Any purchase at any node must be
financed by a sale of equal value at the same node. The sale could be of other
goods, or it could be of securities. The market for securities, as well as for
goods, must clear at every node in the tree. Finally, the rational
expectations hypothesis becomes much more demanding: Agents are not only aware
of the prices of all goods and securities today, but are also aware of how all
the prices depend on each node in the future.

\subsection{Rational Expectations and Utility Maximization}

Maximizing a utility function already embeds a large dose of rationality. The
utility function is defined over the entire consumption plan. The agents are
also presumed to know the tree of states of nature, and the probabilities of
every branch, and to correctly forecast the prices that will emerge at each
node if they get there. So when an agent stops to buy an apple today, she does
so only after considering what other fruit she will likely have for dinner,
how likely it is she will pass the same shop tomorrow, what else she could
spend the money on the day after tomorrow, and so on.

Not surprisingly, this rational utility maximization has received a great deal
of attention and has been attacked from many different directions. In 2002
Kahneman received a Nobel prize for showing that real people do not have
simple utility functions such as those that are typically assumed in standard
economic theory, but instead have preferences that depend on context and even
on framing, i.e. on the way things are presented \cite{Kahneman00}. The field
of behavioral economics, which Kahneman exemplifies, investigates
psychologically motivated modifications of the rationality and utility
assumptions. There are a range of levels at which one can do this. At the
opposite extreme from rational expectations, one can simply assume that agents
have fixed beliefs, which might or might not correspond to reality
\cite{Brock97,Brock98,Brock99}. At an in-between level one can use a so-called
\textit{noise trader} model in which some agents have fixed beliefs while
others are perfectly rational \cite{Schleifer00,Delong90}. (This can actually
make the task of computing the equilibrium even harder, since the rational
agents have to have perfect models of the noise traders as well as of each
other). At a higher level one can assume that agents are not given the
probabilities of states of nature a priori, but need to learn them
\cite{Sargent93,Timmermann01}.

\subsection{Extensions of Equilibrium}

The rational expectations equilibrium model that we have described so far is
highly idealized, and in the next sections we will present a critique
outlining its problems. The main agenda of current economic theory is to
extend it by modifying or generalizing the assumptions. For example, in 1958
Samuelson extended the finite tree equilibrium model to an infinite number of
time periods, calling it the overlapping generations model, in order to study
intergenerational issues like social security. In 2001 Akerlof, Spence and
Stiglitz received the Nobel prize for their work on \textit{asymmetric
information}, i.e. for studying situations such as buying a used car, or
negotiations between labor and management, in which the information sets of
agents differ \cite{Akerlof00,Spence74,Stiglitz81}. Another significant
extension is to allow agents who sell securities to default on delivering the
dividends those securities promise \cite{Dubey05,Geanakoplos97}. This raises
interesting problems such as how the market determines the amount of
collateral that must be put up for loans \cite{Geanakoplos03}.

In view of all these extensions the meaning of the word \textquotedblleft
equilibrium" is not always clear. One definition would be any model that could
be interpreted as satisfying hypotheses (1 - 4) as described in the
introduction to this Section. But we also wish to allow for some boundedly
rational agents. Following the practice of most economists, we will define an
\textit{equilibrium model} as one in which at least some agents maximize
preferences and incorporate expectations in a self-consistent manner. So, for
example, a noise trader model is an equilibrium model as long as some of the
agents are rational, but models in which all the agents act according to fixed
beliefs are not equilibrium models.



\section{Efficiency\label{efficiency}}

In equilibrium everyone is acting in his own selfish interest, guided only by
the market prices, without any direction from a central planner. One wonders
if anything coherent can come from this decentralization of control. One might
expect coordination failures and other problems to arise. If activities were
organized by rational cooperation, surely people could be made better off. Yet
in the general equilibrium model of Arrow and Debreu, equilibrium allocations
are always allocatively efficient. \textit{Allocative efficiency} means that
the economy is as productive as possible, or \textit{Pareto efficient}, i.e.
that there is no change in production choices and trades that would make
everyone better off. In equilibrium everyone acts in his own selfish interest,
yet the decisions they make are in the common good in the sense that not even
a benevolent and wise dictator who told everyone what to produce and what to
trade could make all the agents better off.

In financial equilibrium the situation is much more complicated and
interesting. An important assumption underpinning much of the theory of
financial economics is that of \textit{complete markets}. The markets are
complete if at any node in the tree it is possible to offset any future risk,
defined as a particular future state, by purchasing a security that pays if
and only if that state occurs. When all such securities exist and markets are
complete, financial equilibrium is also Pareto efficient.\footnote{
Brock, Hommes and Wagener have shown that when one deviates from rational
expectations, for example by assuming agents use reinforcement learning,
adding additional securities can destabilize prices \cite{Brock07}. This
suggests that there are situations where market completeness can actually
decrease utility.}

In contrast, when markets are not complete, financial equilibrium is almost
never Pareto efficient. Worse still, the markets that do exist are not used
properly. When markets are incomplete, a benevolent and wise dictator can
almost always make everyone better off simply by taxing and subsidizing the
existing security trades \cite{Geanakoplos86}. What form such government
interventions should take is a matter for economic policy. Is it a good
approximation to assume that real markets are complete? If they are not
complete, does the government have enough information to improve the
situation? These questions have been a matter of considerable debate.


Since allocative efficiency typically fails for financial economies,
economists turned to properties of security prices that must always hold.
\textit{Informational efficiency} is the property that security prices are in
some sense unpredictable. In its strongest form, informational efficiency is
the simple statement that prices are a \textit{martingale}
\cite{Fama70,Fama91}, i.e. that the current price is the best prediction of
future prices.\footnote{
More precisely, if no security pays a dividend in period $t+1$, then there is
a probability measure on the branches out of any node in period $t$ such that
the expected price in period $t+1$ of any security (relative to a fixed
security), given its (relative) price in period $t$, is just the period $t$
price.}
This is important because it suggests that all the information held
in the economy is revealed by the prices. If weather forecasts cannot improve
on today's price in predicting future orange prices, then one can say that
today's orange price already incorporates those forecasts, even if many
traders are unaware of the forecasts.\footnote{
Roll found in some Florida counties that today's orange prices predicted
tomorrow's orange prices better than today's weather reports, and also that
today's orange prices predicted tomorrow's weather better than today's weather
forecasts \cite{Roll84b}.}
\textit{Arbitrage efficiency} means that through
buying and selling securities it is not possible for any trader to make
profits in some state without taking any risk (i.e. without losing money in
some other state). Again this is important because it suggests that there are
no traders who can always beat the market by taking advantage of less informed
traders. Thus the uninformed should not feel afraid that they are not getting
a fair deal. Informational efficiency and arbitrage efficiency are intimately
related; in some contexts they are equivalent.

A consequence of the martingale property is that the price of every
(finitely-lived) asset must be equal to its \textit{fundamental value} (also
called present value if there is no uncertainty). The fundamental value of an
asset is the discounted expected dividend payments of the asset over its
entire life, where the expectation is calculated according to the martingale
probabilities, and the discount is the risk free rate (the interest rate on
securities such as treasury bonds for which the chance of default can be
takent to be zero). This is important for investors who might feel too
ignorant to buy stocks, because it suggests that on average everybody will get
what he pays for.

From the point of view of finance, arbitrage efficiency and informational
efficiency are particularly useful because they can be described without any
explicit assumptions about utility. Many results in finance can be derived
directly from arbitrage efficiency. For example, the Black-Scholes model for
pricing options is perhaps the most famous model in finance. The option price
can be derived from only two assumptions, namely that the price of the
underlying asset is a random walk and that neither the person issuing the
option nor the person buying it can make risk-free profits. Because it does
not rely on utility, several people have suggested that arbitrage efficiency
forms a better basis for financial economics than equilibrium; see for example
Ross \cite{Ross04}. As we will argue later, due to all the problems with
defining utility, this is a highly desirable feature.

\section{The virtues of equilibrium \label{virtues}}

\subsection{Agent Based Modeling}

Equilibrium theory focuses on individual actions and individual choices. By
contrast Marx emphasized class struggles, without asking whether each
individual in a class would have the incentive to carry on the struggle.
Similarly Keynesian macroeconomics often posited reduced form relationships,
such as the positive correlation between unemployment and inflation (called
the Phillips curve), without deriving them from individual actions.
Equilibrium theory is an agent based approach which does not admit any
variables except those that can be explained in terms of individual choices.

The advantages of the agent based approach can be seen in the macroeconomic
correlation between inflation and output. At its face, such a correlation
suggests that the monetary authority can stimulate increased output by
engineering higher inflation through printing money. According to equilibrium
theory, if inflation causes higher output, there must be an agent based
explanation. For example, it may be that workers see their wages going up and
are tempted to work harder, not realizing that the prices of the goods they
will want to buy are also going up so that there is no real incentive to work
harder. But if that is the explanation, then it becomes obvious that policy
makers will not be able to rely on the Phillips curve to stimulate output by
printing more money. Agents will eventually catch on that when their wages are
rising, so are general prices.\footnote{This has come to be called the Lucas
critique \cite{Lucas76}.}
That skepticism was validated during the
stagflation of the 1970s, when there was higher inflation and lower output. If
printing money sometimes causes higher output, it must be through a different
channel, connected say to interest rates and liquidity constraints. Finding
what these are at the agent level deepens the analysis and brings it closer to reality.

\subsection{The need to take human reasoning into account}

The fundamental difference between economics and physics is that atoms don't
think. People are capable of reasoning and of making strategic plans that take
each other's ability to reason into account. Ultimately any economic model has
to address this problem. By going to a logical extreme, equilibrium models
provide a way to incorporate reasoning without having to confront the
messiness of real human behavior. Even though this approach is obviously
unrealistic, the hope is that it may nonetheless be good enough for many purposes.

\subsection{Parsimony}

\subsubsection{Rationality}

The rationality hypothesis is a parsimonious description of the world that
makes strong predictions. Rational expectations equilibrium models have the
advantage that agent expectations are derived from a single, simple and
self-consistent assumption. Without this assumption one needs to confront the
hard task of determining how agents actually think, and how they think about
what others think. This requires formulating a model of cognition or learning,
and thereby introducing additional assumptions that are usually complicated
and/or ad hoc. Without rationality one needs either to introduce a set of
behavioral rules of thumb or to introduce a learning model. While perfect
rationality defines a unique or nearly unique model of the world, there are an
infinite number of boundedly rational models. To paraphrase Christopher Sims,
once we depart from perfect rationality, there are so many possible models it
is easy to become lost in the wilderness of bounded rationality \cite{Sims80}.

Perfect rationality is an impossible standard for any individual to attain,
but the hope of the economist in making a rational expectations model is that
in aggregate, people behave \textquotedblleft as if" they were rational. This
may be true in some situations, and it may not be true in others. In any case,
rational expectations models can provide a useful benchmark for understanding
whether or not people are actually rational, which can serve as a starting
point for more complicated models that take bounded rationality into account.

\subsubsection{Succinctness}

Equilibrium theory provides a unifying framework from which many different
conclusions can be drawn. Rather than having to invent a new method to attack
each problem, it provides a standardized approach, and a standardized language
in which to explain each conclusion.

A standardized model does not rule out new theories. New equilibrium theories
are of two forms. First, one can specialize the class of utilities,
technologies, and endowments to try to draw a sharp and interesting conclusion
that does not hold in general. If either the premise or conclusion can be
empirically validated, one has found a law of economics. Second, one can
extend the definition of equilibrium (adding say asymmetric information or
default). The second form facilitates the discovery of many theories of the
first form. Agreement on a unifying framework, such as equilibrium and
hypotheses (1 - 4), makes it easier to evaluate new economic theories, and to
make progress on applications. But of course it stultifies radically new approaches.


\subsection{The normative purpose of economics}

One of the principal differences between physical and social science is that
the laws of the physical world are fixed, whereas the laws of society are
malleable. As human beings we have the capacity to change the world we live
in. Economics thus has a normative as well as a descriptive purpose. A
\textit{descriptive} model describes the world as it is, while a
\textit{normative} model describes the world as it might be under a change in
social institutions. The equilibrium model is meant to allow us to describe
both. For each arrangement of social institutions, such as those that prevail
today, we compute the equilibrium. The representation of agents by utilities
(as well as endowments) enables us to evaluate the benefits to each person of
the resulting equilibrium, and thereby implicitly the utility of the
underlying institutions. Equilibrium theory then recommends the institutions
that lead to the highest utilities. By contrast, consider a purely
phenomenological model that assumes behavioral rules for the agents without
deriving them from utility. Such a model is either silent on policy questions,
or requires a separate ad hoc notion of desireability in order to recommend
one institution over another.

The rationality hypothesis allows us to define equilibrium succinctly. But as
a byproduct it shapes the economist's evaluation of the institutions. For
example, one of the basic questions in economics is whether free markets
organize efficient sharing of risks, leading to sensible production decisions.
The equilibrium model is often used to show that under certain conditions,
free market incentives lead selfish individuals to make wise social decisions.
We called this allocative efficiency or Pareto efficiency. But of course the
rationality of all the individuals is an indispensable hypthesis in the proof.
If agents are misinformed about future production possibilities, or act
whimsically, the free market economy will obviously make bad decisions.

Even if rational expectations doesn't provide a good model of the present, it
may sometimes provide a good model of the future. The introduction of an
equilibrium model can change the future for the simple reason that once people
better understand their optimal strategies they may alter their behavior. A
good example of this is the Black-Scholes model; as people began using it to
buy and sell mispriced options, the prices of options more closely matched its
predictions. Another example is the capital asset pricing model. In that model
the optimal strategy for every investor is to hold a giant mutual fund of all
stocks. Nowadays every investor who has learned a little finance does indeed
think first of holding exactly such an index (though few investors hold only that).

Unfortunately, many economists have used the normative purpose of economics as
an excuse to construct economic theories that are so far removed from reality
that they are useless. For an economic theory to have useful normative value
it must be sufficiently close to reality to inspire confidence that its
conclusions give useful advice. Unless a theory can give some approximation of
the world as it is, it is hard to have confidence in its predictions of how
the world might be.

\subsection{Why does Wall Street like the equilibrium model?}

Many Wall Street professionals try to beat the market by actively looking for
arbitrages. Fundamental analysts, for example, believe that they can find
stocks whose prices differ substantially from their fundamental values.
Statistical arbitrageurs believe they can find information in previous stock
movements that will help them predict the relative direction of future stock
movements. The activities of these Wall Street professionals seems proof that
they think the equilibrium model is flawed. If they believe the market is in
equilibrium, why are they pursuing strategies that cannot succeed in equilibrium?

Ironically, in fixed income and derivative asset markets the arbitrageurs
themselves use the apparatus of equilibrium models to find arbitrages. Many
people make good profits by betting that when prices make large deviations
from equilibrium these deviations will eventually die out and return to equilibrium.

There are many ways Wall Street practitioners use equilibrium models. We
review some of them below:

\subsubsection{Conditional forecasting is good discipline}

Many pundits, and even some professional economists (and especially
macroeconomists), make unconditional forecasts. They say growth next quarter
will be slow, but by the fourth quarter it should pick up and unemployment
should start to decline. They do not often bother to say, for example, that
their predictions might change if there is a messy war in Iraq. In the
equilibrium model agents do not know what state will prevail tomorrow, but
they do know what prices will prevail conditional on tomorrow's state. They
make conditional forecasts. To do this they construct a tree of possible
states and compute the equilibrium in each state.
Wall Street traders are trained in business schools and economics departments
across the country to appreciate equilibrium models. Nowadays so many Wall
Street traders make conditional forecasts using concrete trees of future
possibilities that this assumption has become more plausible as a descriptive
model of the world.

One of the virtues of making conditional forecasts is that it stimulates
traders to find strategies (called riskless arbitrages) that will work no
matter what the future holds. If one is lucky enough to find a riskless
arbitrage, success is independent of the probabilities of the future states of
nature. In the real world riskless arbitrages are rare (as equilibrium theory
says they should be). Most real arbitrageurs actually make their money with
\emph{risky} arbitrages, and in this case the probabilities for branches of
the tree become important. The estimation errors that are inherent in
estimating these probabilities compound the risk. Not only might the future
bring a state in which money will be lost, but the trader might also
underestimate the probability of this state.\ But at least she can see the
scenarios that constitute the model explicitly, and can compute the
sensitivity of her expected profits to variations in the probabilities of
these scenarios.

\subsubsection{The power of the no-arbitrage hypothesis}

Tree building is most useful when the future possibilities are easy to
imagine, when they are not too numerous to compute, and when there is a
reliable guide to their probabilities. The hypothesis of no-arbitrage
drastically reduces the number of states, making the computation feasible, and
often even determines the probabilities.

One reason that financial practitioners like the no arbitrage assumption is
because it makes things much simpler. Arbitrage efficiency requires a
consistency between prices that drastically reduces the size of the
hypothetical tree. For example, a mortgage derivative trader might worry about
future interest rates of all maturities, including the overnight (i.e one day)
rate, the yearly rate, the ten year rate, and so on. If there are $20$ such
rates, and each rate can take on $100$ values, then each node in the tree will
require $100^{20}$ branches! Using the overnight rate alone requires a much
smaller tree with only $100$ branches from each node. The crucial point is
that if the trader assumes there will never be any arbitrage among interest
rate securities of different maturities in the future, then the smaller tree
will be sufficient to recover all the information in the larger tree even for
a trader who cares about all the rates. Given the smaller tree and the
probabilities of each branch, the trader can deduce all the other future long
rates, conditional on the future overnight rates, without adding any new nodes
to her tree. (Today's two day discount, for example, can be deduced as the
product of today's one day discount and tomorrow's one day discount, averaged
over all possible values of the one day discount tomorrow). Furthermore, if
the trader assumes there is no arbitrage between current long maturity
interest rate securities and future short interest rate securities and current
interest rate options, then she can deduce the probabilities of each branch in
the smaller overnight interest rate tree.

The homage arbitrage traders pay to arbitrage efficiency is the most
compelling evidence that in some respects it is true. Again we see the irony
that the road to arbitrage comes from assuming no arbitrage.\footnote{
The mortgage trader assumes no arbitrage in the class of interest rate
securities alone. By contrast, she does assume there is an arbitrage involving
interest rate securities and mortgage securities. A mortgage derivative trader
might assume there is no arbitrage involving interest rate securities and
mortgage securities, but that there is an arbitrage once mortgage derivatives
are added to the mix.}

\subsubsection{Risk reduction}

\textit{Hedging} refers to the process of reducing specific financial risks.
To hedge the value of a particular security one creates a combination of other
securities, called a \textit{portfolio}, that mimics the security that one
wishes to hedge. By selling this combination of securities one can cancel or
partially cancel the risk of buying the original security. Statistical
arbitrageurs, for example, try to make bets on the relative movements of
stocks without making bets on the overall movement of the stock market. The
market risk can be hedged by making sure that the total position is market
neutral, i.e. that under a movement of the market as a whole the value of the
position is unaffected. This is typically done by maintaining constraints on
the overall position, but can also be done by selling a security, such as a
futures contract on a market index.

To properly hedge it is necessary to have a set of scenarios for the future.
Such scenarios can be represented by means of a tree, exactly as we described
for equilibrium models. Thus, the same structure we defined in order to
discuss equilibrium is central to hedging risks. Wall Street professionals are
very concerned about risks, and for this reason like the financial framework
for understanding equilibrium.

Hedging and arbitrage are two sides of the same coin. In both hedging and
arbitrage, the trader looks for a portfolio of securities that will pay off
exactly what the risky asset does. By buying the asset and selling the
portfolio, the trader is completely hedged. If the trader discovers that the
cost of the hedging portfolio is more than the security, then hedging becomes
arbitrage: The arbitrageur can buy the security and sell the portfolio and
lock in a riskless profit.

\section{Difficulties and limitations of equilibrium\label{limitations}}

While equilibrium should be an important component of the economist's tool
kit, its dangers and limitations need to be understood (and all too often are not).

\subsection{Falsifiability?}

Empirical laws in economics are much harder to find than in physics or the
other natural sciences. Realistic experiments for the economy as a whole are
nearly impossible to conduct (though for some small scale phenomena there is a
thriving experimental research effort underway) and many causal variables are
impossible to measure or even to observe. Economists usually do not dream of
finding \textquotedblleft constants of behavior," analogous to the constants
of nature found in physics; when they do, they do not expect more than a
couple of significant digits.\footnote{
Okun's Law for example, which states that every 1\% increase in unemployment
is accompanied by a 3\% decrease in GNP, has one significant digit.} We shall
argue later that this may be a mistake, and that it is possible to find finely
calibrated relationships in economic data.

Perhaps because of the difficulty of empirical work, economic theory has
emphasized understanding over predictability. For example, to economists,
equilibrium theory itself is most important not for any empirical predictions,
but for the paradoxical understanding it provides for markets: Without any
centralized coordination, all individual plans can perfectly mesh; though
everyone is perfectly selfish, their actions promote the common good; though
everybody is spending lots of time haggling and negotiating over prices, their
behavior can be understood as if everybody took all the prices as given and immutable.

Equilibrium theory has not been built with an eye exclusively focused on
testability or on finding exact functional forms. Most equilibrium models are
highly simplified at the sacrifice of features such as temporal dynamics or
the complexities of institutional structure. When it comes time to test the
model, it is usually necessary to make auxiliary assumptions that put in
additional features that are outside of the theory. All these factors make
equilibrium theories difficult to test, and mean that many of the predictions
are not as sharp as they seem at first. In many situations, whether or not the
predictions of equilibrium theory agree with reality remains controversial,
even after decades of debate.

Not only are there few sharp economic predictions to test, but the hypotheses
of economic equilibrium also seem questionable, or hard to observe. A
long-standing dispute is whether utility is a reasonable foundation for
economic theory. The concept of utility as it is normally used in economic
theory is purely qualitative. The functional form of utility is generally
chosen for convenience, without any empirical justification for choosing one
form over another. No one takes the functional form and the parameters of
utility functions literally. This creates a vagueness in economic theory that
remains in its predictions.

Psychologists have generally concluded that utility is not a good way to
describe preferences, and have proposed alternatives, such as prospect theory
\cite{Kahneman00}. Some economists have taken this seriously, as evidenced by
the recent Nobel prize awarded to Kahneman. However, most theory is still
built around conventional utility functions, and so far the alternatives are
not well-integrated into the mainstream. Attempts that have been made to
develop economic models based on prospect theory still do not determine the
parameters \textit{a priori} (and in fact they have more free parameters).
Thus so far it is still not clear whether more general notions of preferences
can be used to make sharper and more quantitative economic
predictions.\footnote{
One approach that has gained some attention is hyperbolic discounting
\cite{Berns07}. Hyperbolic agents care a lot more about today than tomorrow,
yet they act today as if they will never care about the difference between 30
days from today and 31 days from today.}

The other big problem that is intrinsic to equilibrium theories concerns
expectations about the probabilities of future states. It is difficult to
measure expectations. In practice we only observe a single path through the
tree of future states. The particular path that is observed historically may
be atypical, and may not be a good indication of what agents really believed
when they made their decisions. Thus, even when we have a good historical
record there is plenty of room to debate the conclusions.

All these problems occur in testing the assertion that markets are arbitrage
efficient, as we discuss later. Real arbitrages are rarely risk free. Instead,
they involve risk, and determining whether one arbitrage is better than
another involves measuring risk. Future risks may not match historical risks.
If a skillful trader produces excess returns (above and beyond the market as a
whole) with a high level of statistical significance, a champion of market
efficiency can always argue that this was possible only because the asset had
\textquotedblleft unobserved risk factors\textquotedblright. What seems like a
powerful scientific prediction from one point of view can look suspiciously
like an unfalsifiable belief system from another.

\subsection{Parsimony and tractability are not the same}

While we have argued that equilibrium models are parsimonious this does not
necessarily mean that they are easy to use. The mathematical machinery
required to set up and solve an equilibrium model can be cumbersome. Finding
fixed points is much more difficult than iterating dynamic maps. Thus
incorporating equilibrium into a model comes at a large cost, and may force
one to neglect other factors, such as the real structure of market
institutions. As a result of this complexity most equilibrium models end up
being qualitative toy models, formulated over one or two time periods, whose
predictions are too qualitative to be testable.

\subsection{Realism of the rationality hypothesis? \label{realism}}

It seems completely obvious that people are not rational. The economic model
of rationality not only requires them to be super-smart, it requires them to
have God-like powers of omniscience. This should make anyone skeptical that
such models can ever describe the real world. Rational expectations (i.e.
knowing the probabilities of each branch in the tree and knowing what the
prices will be at each node) is often justified by the argument that people
behave \textquotedblleft as if" they were rational. Skeptics do not find such
arguments convincing, particularly without supporting empirical evidence.
There are many reasons to be suspicious of equilibrium models, and many
situations where the cognitive tasks involved are sufficiently complicated
that the a priori expectation that an equilibrium theory should work is not high.

The conceptual problems with perfect rationality can be broken down into
several categories:

\begin{itemize}
\item \textit{Omniscience.} To take each others' expectations into account
agents must have an accurate model of each other, including the cognitive
abilities, utility, and information sets of all agents. All agents construct
and solve the same tree. They must also agree on the probabilities of the
branches. More realistically one must allow for errors in model building, so
that not all agents have the same estimates or even the same tree of possibilities.

\item \textit{Excessive cognitive demands}. The cognitive demands the
equilibrium model places on its agents can be preposterous in the sense that
the calculations the agents need to make are extremely time consuming to
perform. Even given the tree and the probabilities and the conditional prices,
it may be difficult for an agent to compute her optimal plan. But how does the
agent know the conditional prices unless she herself computes the entire
equilibrium based on her knowledge of all the other agents' utilities and
endowments? These computational problems can be intractable even if all agents
are fully rational, and they can become even worse if some agents are rational
and others aren't.

\item \textit{Behavioral anomalies}. There is pervasive evidence of
irrationality in both psychological experiments and real world economic
behavior. Even in simple situations where people should be able to deal with
the difficulties of computing an equilibrium it seems many do not do so
\cite{Barberis03,Thaler05}.

\item \textit{Modeling cognitive cost and heterogeneity.} Real agents have
highly diverse and context dependent notions of rationality. Because models
are expensive to create real agents take shortcuts and ignore lots of
information. They use specialized strategies. The resulting set of decisions
may be far from the rationality supposed in equilibrium, even when taken in aggregate.
\end{itemize}

\subsection{Limited scope\label{limitedScope}}

There are many interesting problems in financial economics that the
equilibrium framework was never intended to solve. Under conditions that are
rapidly evolving, e.g. where there is insufficient time for agents to learn
good models of the situation they are placed in, there is little reason to
believe that the equilibrium framework describes the real world. Some of the
problems include

\begin{itemize}
\item \textit{Evolution of knowledge.} In a real market setting the framework
from which we view the world is constantly and unpredictably evolving. The
models used by traders to evaluate mortgage securities in the 1980s were
vastly more primitive than the models used to evaluate the same securities
today. People have a difficult time imagining the possible ways in which
knowledge will evolve, much less assigning probabilities to all of its states.

\item \textit{Lack of tatonnement.} The fact that an equilibrium exists does
not necessarily imply that it is stable. For stability it is necessary to show
that when a system starts out of equilibrium it will necessarily move toward
equilibrium. This requires the construction of a model for price formation out
of equilibrium, a process that is called tatonnement. Such models suggest that
there are many situations where prices will fail to converge to equilibrium
\cite{Fisher83}.

\item \textit{Inability to model deviations from itself}. Equilibrium can't
model deviations from itself. It provides no way to pose questions such as
``How efficient is the economy?''. One would like to be able to understand
questions such as the timescale for violations of arbitrage efficiency to
disappear, but such questions are inherently outside of the equilibrium
framework. (See the discussion in Section~\ref{motivation}).
\end{itemize}

While the financial implications of equilibrium theory are on one hand
extraordinarily powerful, on the other hand, their very power makes them
almost empty. Economic theory says that there is very little to know about
markets: \ An asset's price is the best possible measure of its fundamental
value, and the best predictor of future prices. If that is true, there is no
need to answer the kinds of questions that economists are asked all the time,
such as \textquotedblleft Are stocks going up or down?\textquotedblright, or
\textquotedblleft Is this a good time to invest in real
estate?\textquotedblright. Only by going outside the equilibrium model can one
raise many pressing financial questions.

It is worth noting that the mere fact that people so persistently ask such
questions puts equilibrium theory into doubt.\ There are three possibilities:
\ Either the people that do this are irrational, but if they are, then how can
we expect equilibrium theory to describe them? \ Or they are rational, but if
so, then their questions reflect that their understanding is deeper than that
of the equilibrium theorist. The only other explanation is that the people who
ask such questions do not invest an economically significant amount of money.
\ But this seems implausible -- almost everyone asks such questions.

\section{Empirical evidence for and against \label{empiricalEvidence}}

Equilibrium theory teaches us that expectations are important, and this is no
less true in evaluating the success or failure of equilibrium theory itself.
The booster can point to significant successes and the detractor can point to
significant failures.

The booster can assert that equilibrium theory has made many practical
suggestions that have been followed and have been borne out as correct.
Communism did not produce as much output as the free markets of capitalism.
Diversifying your investments really is a good idea, and many index funds have
been created to accommodate this desire.
A hedge fund that can show its returns are independent of much of the rest of
the economy can attract investors even if it promises a lower rate of return
than competitors whose returns are highly correlated with the
market.\footnote{
Though diversification is often sighted as evidence for equilibrium theory,
the skeptic will point out that this only depends on portfolio theory, which
is a simple result from variational calculus concerning statistical estimation
and has nothing to do with equilibrium theory.}
Arbitrages are indeed hard to
find, and even harder to exploit. Stock option prices are explained to a high
degree of accuracy by the Black-Scholes formula. Prepayment behavior for prime
mortgages can be explained by maximizing the utility of individual households
(with some allowances for inattention and other irrational behaviors).

The skeptic counters that, with a few exceptions, most of the predictions of
equilibrium theory are qualitative and have not been strongly borne out
empirically in an unambiguous manner. To quote Ijiri and Simon:
\textquotedblleft To be sure, economics has evolved a highly sophisticated
body of mathematical laws, but for the most part these laws bear a rather
distant relation to empirical phenomena, and usually imply only qualitative
relations among observables ... Thus, we know a great deal about the direction
of movement of one variable with the movement of another, a little about the
magnitudes of such movement, and almost nothing about the functional forms of
the underlying relations\textquotedblright\ \cite{Ijiri77}. Since they said
this in 1977 the situation has improved, but only a little. Apart from options
and prime mortgage pricing (but not subprime), there are very few examples of
economic theories that cleanly fit the data and also unambiguously exclude
nonequilibrium alternatives.
While some of the predictions of equilibrium models are in qualitative
agreement with the data, there are few that can be quantitatively verified in
a really convincing manner.

One of the conclusions of equilibrum theory is that relative prices should
follow a martingale. It is well documented that over the long term risky
securities like stocks have had higher returns than safe securities like short
term government bonds, at least in the United States.\footnote{Brown, Goetzmann and
Ross point out that averaged over all western countries, stocks have not
outperformed bonds \cite{Brown95}. An economic collapse, say during the Russian revolution,
usually crushes stock prices more than bond prices. So the American experience
may not be representative. Of course this illustrates the difficulty of
testing the prediction.} This would seem to contradict the martingale pricing
theorem. But the martingale pricing theorem says only that asset prices follow
a martingale with respect to some probability measure, not necessarily with
respect to the objective probabilities. The skeptic naturally complains that
this is giving the theory too much freedom, making it close to tautological.
Economists have responded by making more assumptions on the underlying
utilities in the equilibrium model to limit how the set of allowable
martingale probabilities can deviate from objective probabilities, making the
theory falsifiable. The most famous model of that type is the so-called
capital asset pricing model (CAPM) developed by Markowitz, Tobin, Linter, and
Sharpe \cite{Markowitz52,Tobin58,Lintner65,Sharpe64}. In one version of CAPM all utilities are quadratic.\footnote{This is
an example of the first kind of progress equilibrium theory makes, in which
specialized assumptions give rise to more precise conclusions, as we discussed
in Section 4.3.2.} Under this assumption the martingale probabilities are not
arbitrary, but must lie in a one dimensional set, making the theory more
straightforwardly testable. CAPM does explain why riskier securities should
have higher returns than safe securities, and also provides a rigorous
definition and quantification of the risk of a security. Data from the 1930s
through the 1960s seemed to confirm the CAPM theory of pricing. But little by
little the empirical validity of CAPM has unraveled. Of course there is always
the possibility that a different specialization will give rise to another
model that matches the data more closely. Or perhaps an extension of the
equilibrium model involving default and collateral will be able to match the
data, without creating so many free parameters as to make the theory
unfalsifiable.\footnote{If successful, this would be an example of the second
kind of progress equilibrium theory makes, when the model is extended to
incorporate new concepts, but retains methodological premises 1-4, as we
discussed in Section 4.3.2.}

A similar story can be told about the conclusion that asset prices should
match fundamental values. Attempts to independently compute fundamental values
based on dividend series do not match prices very well. For example, for the
U.S. stock market Campbell and Shiller~\cite{Campbell89} show deviations
between prices and fundamental values of more than a factor of two over
periods as long as decades. This conclusion can be disputed on the grounds
that fundamental values are not easily measurable, and perhaps none of the
measures they use are correct (though they tried several alternatives). But
whether the theory is wrong or whether it is merely unfalsifiable, the failure
to produce a good match to fundamental values represents a major flaw in
either case.

Perhaps even worse, there are situations where equilibrium makes predictions
that appear to be false. For example, people seem to trade much more than they
should in equilibrium \cite{Sebenius83}. Global trading in financial markets
is on the order of a hundred times as large as global production
\cite{Shiller81,Shiller97}. If people are properly taking each others'
expectations into account why should they need to trade so much? One of the
problems is that the theoretical predictions of how much people should trade
are not very sharp, due at least in part to the problems discussed in the
previous section. Nonetheless, it seems that even under the most favorable set
of assumptions people trade far more than can be rationalized by an
equilibrium model~\cite{Odean99}.

As already discussed in the previous section, one of the key principles of
rational expectations is that investors should correctly process information,
and thus price movements should occur only in response to new information.
Studies that attempt to correlate news arrival with large market moves do not
support this. For example, Cutler, Poterba and Summers examined the largest
$100$ daily price movements in the S\&P index during a $40$ year period and
showed that most of the largest movements occur on days where there is no
discernable news, and conversely, days that are particularly newsworthy are
not particularly likely to have large price movements \cite{Cutler89}. Other
studies have shown that price volatility when markets are closed, even on
non-holidays, is much lower than when the market is open \cite{French86}. The
evidence seems to suggest that a substantial fraction of price changes are
driven by factors unrelated to information arrival. While news arrival and
market movements are clearly correlated, the correlation is not nearly as
strong as one would expect based on rational expectations. As asserted in the
previous section, markets appear to make their own news. One of the problems
in answering this question empirically is that it is difficult to measure
information arrival-- \textquotedblleft news" contains a judgement about what
is important or not important, and is difficult to objectively reduce to a
quantitative measure.\footnote{
Engle and Rangel \cite{Engle05} demonstrate that
there is a positive correlation between low frequency price volatility and
macroeconomic fundamentals. Because their results involve both a longtitudinal
and a cross-sectional regression involving both developed and undeveloped
countries the interpretation of the correlations that they observe are not
obvious. No one disputes that prices respond to information -- the question
is, \textquotedblleft What fraction of price movements can be explained by
information?". Timescale is clearly critical, i.e. one expects more
correlation to information arrival at lower frequencies.}

\section{Motivation for non-equilibrium models: a few
examples\label{motivation}}

We begin by describing some empirical regularities in financial data that seem
salient yet which equilibrium theory does not seem able to explain. Of
particular importance is price volatility. Not only do many people think that
prices change more than they should under equilibrium, but also volatility
displays an interesting temporal correlation structure that seems to be better
explained by nonequilibrium models. This correlation is an example of a power
law, a functional form that seems to underly many of the regularities in
financial economics. In physics power laws can't be generated at equilibrium
and are a signature of nonequilibrium behavior. Economic and physical
equilibrium are quite different, however, as we try to make clear. Statistical
testing for power laws is difficult and the existence of power laws and their
relevance for economics have generated a great deal of debate, which we
briefly review.

Next we observe that one of the most frequently cited pieces of evidence for
equilibrium is that arbitrage opportunities tend to disappear. There is
substantial evidence that real markets are efficient at some level of
approximation. But we present anecdotal evidence that efficiencies disappear
surprisingly slowly. We would like a theory of the transition to equilibrium,
or of tatonnement as it has sometimes been called. We end this section with a
brief discussion of one of the most remarkable episodes in financial markets
of the last ten years, which illustrates both the value and the limitations of
the equilibrium model.

\subsection{Some economically important phenomena may not depend on
equilibrium\label{mayNotDepend}}

There are many empirically observed properties of markets that have so far not
been explained under the equilibrium framework. One famous example is called
\textit{clustered volatility}. This refers to the fact that there are
substantial and strongly temporally correlated changes in the size of price
movements at different points in time \cite{Engle82}. If prices move a lot on
a Tuesday, they will probably move a lot the Wednesday after, and probably
(but not as likely) move a lot on Thursday.\footnote{
Mandelbrot and Clark
suggest that it is possible to describe clustered volatitlity as if
time gradually speeds up and slows down, according to some random process that
is independent of the direction of prices \cite{Mandelbrot63,Mandelbrot73,Clark73}.  It has also been suggested that the correct notion of trading time can be measured by either transaction volume or frequency of transactions \cite{Ane00}; more careful analysis reveals that this is not correct, and that fluctuations of volume are dominated by fluctuations in liquidity \cite{Gillemot05}.}.
The hard core rational
expectations booster would say that there is nothing to explain. If the
standard equilibrium theory is correct, changes in price are caused by the
receipt of new information, so clustered volatility is simply a reflection of
non-uniform information arrival. The rate of information arrival has to do
with meteorology, sociology, political science, etc. It is exogenous to
financial economics and so it is someone else's job to explain it, if such a
question is even interesting.

An alternative point of view is that, due to a lack of a perfect rationality,
the market is not in perfect equilibrium and thus prices do not simply
passively reflect new information. Under this view the market acts as a
nonlinear dynamical system: agents process information via decision-making
rules that respond to prices and other inputs, and prices are formed as a
result of agent decisions. Since this information processing is imperfect, the
resulting feedback loop amplifies noise. As a result a significant component
of volatility is generated by the market itself. This point of view is
supported by the fact that there are now many examples of non-equilibrium
models that generate clustered volatility, and in some cases there is a good
match to empirical data. For many of these models it is possible to include
equilibrium as a special case; when this is done the overall volatility level
drops and clustered volatility disappears (see
Section~\ref{heterogeneousAgents}). Another relevant point is that volatility
is widely believed to exhibit long-memory, which we discuss shortly in the
section on power laws.

Since the discussion of properties of markets that are not currently explained
by equilibrium makes more sense in the context of alternative models, we will
return to review this in more depth when we discuss non-equilibirum
heterogeneous agent models in Section~\ref{heterogeneousAgents}. As we discuss
in the next section, the questions of when equilibrium theory is irrelevant or
simply wrong can blur together.

\subsection{Power laws\label{powerLaws}}

As we already mentioned, clustered volatility and many other phenomena in
economics are widely believed to display power laws. Loosely speaking, a power
law is a relationship of the form $y=kx^{\alpha}$, where $k$ and $\alpha$ are
constants, that holds for asymptotic values of a variable $x$, such as
$x\rightarrow\infty$ \cite{Newman04,Farmer05b}. The first observation of a
power law (in any field) was made by Pareto, who claimed that this functional
form fit the distribution of the incomes of the wealthiest people in many
different countries \cite{Pareto96}. Power laws are reported to occur in many
aspects of financial economics, such as the distribution of large price
returns\footnote{
Under time varying volatility the distribution of returns can be interpreted
as a mixture of normal distributions with varying standard deviations. This
generically fattens the tails, and for the right distribution of standard
deviations can produce a power law.}
\cite{Mandelbrot63,Fama65,Officer72,Akgiray89,Koedijk90,Loretan94,Mantegna95,Longin96,Lux96,Muller98,Plerou99,Rachev00,Goldstein04b}%
, the size of transactions \cite{Gopikrishnan00,Lillo05b}, the autocorrelation
of volatility \cite{Ding93,Breidt93,Harvey93,Baillie96,Bollerslev96,Lobato00}
the prices where orders are placed relative to the best prices
\cite{Bouchaud02,Zovko02,Mike05}, the autocorrelation of signs of trading
orders \cite{Bouchaud04,Bouchaud04b,Lillo03c,Farmer06}, the growth rate of
companies \cite{Stanley96,Amaral97,Plerou99b}, and the scaling of the price
impact of trading with market capitalization \cite{Lillo03d}. The empirical
evidence for power laws in economics and their relevance for economic theory
remains controversial \cite{Durlauf03}. But there is a large literature
claiming that power laws exist in economics and unless one is willing to cast
aside this entire literature, it seems a serious problem that there is no
explanation based on equilibrium.

This problem might be resolved in one of several ways: Either as in physics,
(1) power laws represent nonequilibrium phenomena from an economic point of
view and thus provide a motivation for nonequilibrium theory, or (2) power
laws are consistent with economic equilibrium, but still need to be explained
by some other means, or (3) equilibrium is a key concept in understanding
power laws but we just don't know how to do this yet. Or perhaps the existence
of power laws is just an illusion. But it seems that power laws fit the data
to at least a reasonable degree of approximation. Furthermore, there are many
non-equilibrium models that naturally generate power laws, and for such models
it is easy to test that the phenomena they generate are indeed real power laws.

It may seem strange to single out a particular functional form for special
attention. There are several reasons for viewing power laws as important. One
of them comes from extreme value theory \cite{Embrechts97}, which says that in
a certain sense there are only four possible convergent forms for the tail of
a probability distribution\footnote{
The four possible tail behaviors of probability distributions correspond to
distributions with finite support; thin tailed distributions such as
Gaussians, exponentials or log-normals; distributions where there is a
critical cutoff above which moments don't exist (namely power laws); and
distributions that lack any regular tail behavior at all. Almost all commonly
used distributions are in one of the first three categories.},
and power laws
are one of them. The main motivation for looking at tail distributions is that
they provide a clue to underlying mechanisms: different mechanisms or
behaviors tend to generate different classes of tail distributions.

Statistically classifying a tail distribution is inherently difficult because
one does not know a priori where the tail begins. In extreme value theory a
power law is more precisely stated as a relation of the form $y=K(x)x^{\alpha
}$, where $K(x)$ is a slowly varying function. A slowly varying function
$K(x)$ satisfies $\lim_{x\rightarrow\infty}K(tx)/K(x)=1$, for every positive
constant $t.$ There is a great variety of possible slowly varying functions;
with a finite amount of data it is only possible to probe to a finite value of
$x$ and the slowly varying function may converge slowly. In practice one tends
to look for a threshhold $x$ beyond which the points $(\log x,\log y)$ almost
lie on a straight line. (This can be made precise using maximum likelihood
estimation, including the estimation of the threshold $x$ \cite{Clauset07}).

In the last twenty years or so the phenomenon of power laws has received a
great deal of attention in physics. Power laws are special because they are
self-similar, i.e. under a scale change of the dependent variable $x^{\prime
}=cx$, treating $K$ as a constant, the function $f(x^{\prime}) = c^{\alpha
}f(x)$ remains a power law with the same exponent $\alpha$ but a modified
scale $K^{\prime} = c^{\alpha}K$. One of the places where power laws are
widespread is at phase transitions, such as the point where a liquid changes
into a solid. Self-similarity proved to be a powerful clue to understanding
the physics of phase transitions and led to the development of the
renormalization group, which made it possible to compute the properties of
phase transitions precisely. The main assumption of the renormalization group
is that the physics is invariant across different scales, i.e. the reason the
observed phenomena are power laws is because their underlying physics is
self-similar. This is an important clue about mechanism.

In physics power laws are associated with nonequilibrium behavior and are
viewed as a possible signature of non-equilibrium dynamics
\cite{Newman04,Farmer05b,Farmer05c}. It was shown more than 100 years ago by
Boltzmann and Gibbs that in physical equilibrium, energies obey an exponential
distribution; thus a power law indicates nonequilibrium behavior. As discussed
in Section~\ref{physicsEq}, equilibrium in physics is quite different from
equilibrium in economics, but it seems telling that the current set of
financial models in which power laws appear are (economic) nonequilibrium
models, as discussed in Section~\ref{heterogeneousAgents}.

\subsection{The progression toward market
efficiency\label{efficiencyProgression}}

\begin{figure}[ptb]
\begin{center}
\includegraphics[scale=0.4]{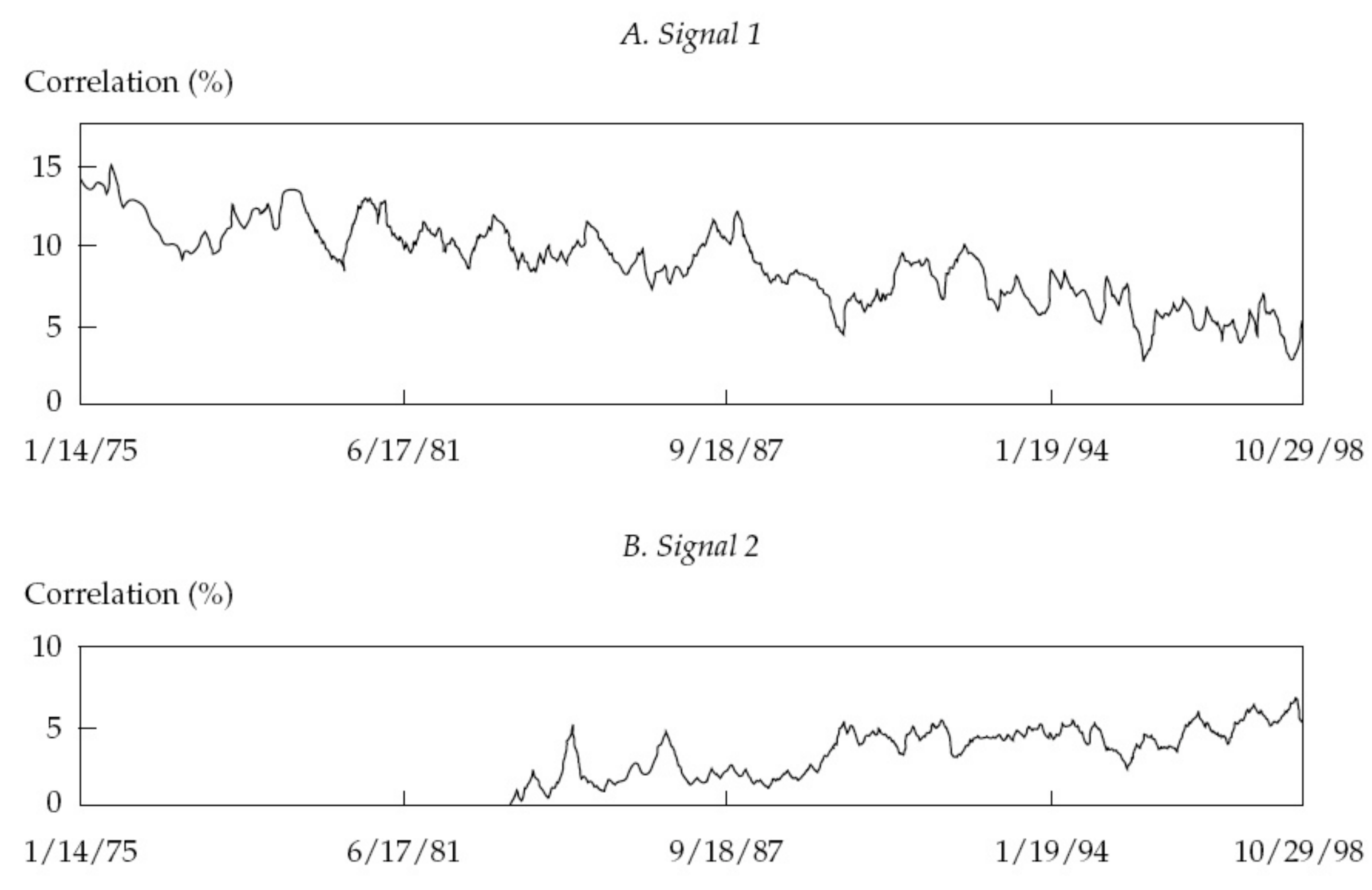}
\end{center}
\caption{The inefficiency of the market with respect to two financial
strategies over a 23 year period. The y axis plots the historical performance
for a backtest of two trading signals developed by Prediction Company.
Performance is measured in terms of the correlation of each signal to the
movement of stock prices two weeks in the future.}%
\label{efficiencyFigure}%
\end{figure}

The theory of market (arbitrage) efficiency is justified by the assertion that
if there were profit making opportunities in markets, they would be quickly
found and exploited, and the resulting trading activity would change prices in
ways that would remove them. But is this really true? Anecdotal empirical
evidence suggests otherwise, as illustrated in Figure~\ref{efficiencyFigure},
where we plot the performance of two proprietary predictive
trading signals developed by Prediction Company (where one of the authors was
employed). The performance of the signals is measured in terms of their
correlation to future price movements on a two week time scale over a twenty
three year period\footnote{
The signals were developed based on data from 1991 - 1998. Afterward data from
1975 - 1990 was acquired and the model was tested on this data without
alteration. Thus the earlier data is completely out of sample and the enhanced
performance prior to 1991 cannot be due to overfitting.}. In the case of
signal 1 we see a gradual degradation in performance over the twenty three
year period. The correlation to future prices is variable but the overall
trend is clear. At the beginning of the period in 1975 it averages about
$14\%$ and by 1998 it has declined to roughly $4\%$. Nonetheless, even in 1998
the signal remains strong enough to be profitably tradeable.

Does this evidence support market efficiency? On one hand the answer is yes:
The usefulness of the signal declines with time. On the other hand it is no:
It takes at least 23 years to do this, and at the end of the period the
opportunity for profits remains present, even if greatly diminished. The
performance of signal 2 is even more surprising, and clearly does not support
market efficiency. Due to a structural change in the market, signal 2 begins
in 1983 and subsequently grows in predictive power over the next decade and a
half. This is the opposite of what standard dogma about efficient markets
would suggest.

The need for a non-equilibrium theory is apparent. Equilibrium theory predicts
that markets are perfectly efficient, and thus violations of efficiency cannot
be addressed without going outside it. The time scale for the degradation of a
profitable strategy is an inherently disequilibium phenomenon.

An estimate of the time scale for the progression toward efficiency can be
made by taking into account what is needed to learn a good model and to
acquire the capital to exploit it. Assume that there is a structural change in
the market, suddenly creating a new inefficiency. For example, suppose that
lots of mortgages are pooled together and securitized for the first time. On
account of lack of historical data or irrational fear (say of adverse
selection), the security might sell for a very low price. How long will this
inefficiency take to disappear? Assume also that the only possible strategies
for removing the inefficiency are econometric, i.e. they are based purely on
statistical time series modeling. In this case one can approximate the rate at
which the strategy can be discovered and exploited based on its Sharpe ratio
$S$, which is defined as the ratio of the expected excess return of the
strategy compared to its risk as measured by the standard deviation of its
profits. Under the optimistic assumption that excess returns are normally
distributed, the statistical significance with which an inefficiency with
Sharpe ratio $S$ can be detected is roughly $S\sqrt{t}$, where $t$ is the time
the inefficiency has been in existence. On an annual time scale a trading
strategy with a Sharpe ratio of one is considered highly desirable\footnote{
The historical Sharpe ratio of the S\&P index is about $0.4$, and the Sharpe
ratio of many famous investors, such as Warren Buffet or George Soros, is
significantly less than one.}. Thus to detect the existence of an inefficiency
at two standard deviations of significance requires four years. Accumulating a
statistically significant track record based on live trading, which is
normally a prerequisite to raising capital, takes another four years. We thus
expect that the typical time scale before such strategies become profitable
should be at least eight years\footnote{There is always the possibility of
conducting more sophisticated research to speed up the learning time. Instead
of observing the returns of the securitized MBS, one could find historical
records of individual loan performance from before the securitization. But
this data is hard to get, and the partial data that can be obtained might not
be selected in the same way the securitized loans were.}. One also has to
address the question of how prices are altered and what is required to make
the inefficiency disappear. As investors start to recognize the opportunity
and exploit it they will make it shrink (say by bidding up the price of the
mortgage backed security) and thus make it harder for others to see, extending
the time until its superior profitability is completely extinguished
\cite{Farmer02}. In any case this argument at least indicates that we should
not be surprised that inefficiencies disappear slowly.

As originally pointed out by Milton Friedman, the idea of a fully efficient
market is inherently contradictory. In order to remove market inefficiencies
we must have traders who are motivated to exploit them. But if the market is
perfectly efficient there is no possibility to make excess profits. While
efficiency might be true at first order, it cannot be true at second order:
There must be on-going violations of efficiency that are sufficiently large to
keep traders motivated. Developing a theory capable of quantitatively
explaining this is a major challenge for economics, and a significant
motivation for developing a non-equilibrium theory.

But what about signal 2 in Figure~\ref{efficiencyFigure}? In this case, the
inefficiency actually \textit{grows} for a period of more than a decade. We
believe this is an example of a structural change. In fact the information
available for constructing the signal was not even available prior to 1983,
and only became available after a rule change in reporting. There are many
examples of sudden structural changes in markets, for example the introduction
of a new derivative such as a mortgage-backed security, a market for a new
type of good such as an internet company, or the introduction of a new
technology such as computers that enables new financial strategies such as
program trading. When the ongoing introduction of financial innovations is
combined with the slow time scale for discovering such signals, perhaps the
persistent existence of inefficiencies is not so surprising.

\subsection{Equilibrium and non-equilibrium in physics and
economics\label{physicsEq}}

To understand the motivations for non-equilibrium models, and to give some
insight into how this might be useful in economics, we make a comparison with
physics. The very name \textquotedblleft non-equilibrium" indicates something
that is defined in terms of its opposite, so before we can discuss
non-equilibrium models in physics we have to make it clear what equilibrium
means and how it differs from the same concept in economics.

As in economics, the word equilibrium in physics has several meanings. In
mechanics the word \textquotedblleft equilibrium" sometimes denotes the fairly
trivial notion that forces are balanced, so there is no acceleration. The more
common use of the term is in thermodynamics or statistical mechanics, where
the word \textquotedblleft thermal equilibrium" means that two systems in
thermal contact reach the same temperature and cease to exchange energy by
heat. A system in equilibrium will experience no change when isolated from its
surroundings, and exists at the minimum of a thermodynamic potential. This is
typically a good approximation only when a system has been left undisturbed
for a sufficiently long period of time.

While the notion of equilibrium in physics is fundamentally different from
that in economics, it does have some similarities. Both involve the
simplifying assumption that the basic set up of the underlying situation is
not changing in time. In economic equilibrium this is implicit in the sense
that in order to imagine that rationality is a reasonable assumption all
agents must have time to make good models of the world. In practice good
models cannot be built without a learning period, in a setting that is
reasonably constant. In physics, equilibrium explicitly requires time
independence. Equilibrium in economics allows explicit time dependences, but
requires that agents have a full understanding of them.

To illustrate how non-equilibrium models are used in physics, suppose you put
a can of cold beer in a warm room. Heat initially flows from the room into the
beer -- the system is out of equilibrium. With time the beer gets warmer, and
eventually to a good approximation the beer comes to room temperature and the
system is at equilibrium.

Equilibrium thermodynamics tells us some useful things, such as the change in
the pressure inside the can of beer between the time it is taken out of the
refrigerator and when it comes to room temperature. There are, however, many
things it doesn't tell us, such as how long will it take for the beer to get
warm. (The analogy to the problem of inefficiencies in the previous section
should now be clear). From an empirical point of view, anyone who drinks beer
knows that on a hot day it will become unpleasantly warm in less than an hour.
A more quantitative, first principles prediction is possible using Fourier's
law, which is a nonequilibrium principle stating that the rate of change in
temperature is proportional to the temperature difference between the beer and
the room. This implies that the temperature of the beer converges to that of
the room exponentially. This can be derived from the heat equation (which is a
standard diffusion equation and would look very familiar to financial
economists). Furthermore, by making more detailed models it is possible to
estimate the parameters of the diffusion process, so that we can roughly
estimate when the beer will become too warm, even if we have no previous
experience drinking beer in warm rooms.\footnote{
This is not at all trivial. The rate of heat transfer depends on the heat
conductivity of the can, on the convection properties of air, and several
other factors. It is nonetheless possible to make estimates based on first
principles.}
Thus we see that in physics, equilibrium theory only explains
part of what we need to know -- to fully understand what is going on we need a
more detailed nonequilibrium theory that can explain the dynamics of the
transitions between equilibrium states. One important advantage of a
nonequilibrium theory is that it makes it possible to know when the
equilibrium theory is valid.

The fact that we have no similar disequilibrium theory in economics is clearly
a serious problem. If there were such a theory, we would know when equilibrium
applies and when it doesn't. There have been many attempts to create such a
theory. Examples can be found in Franklin Fisher's book,
\textit{Disequilibrium foundations of equilibrium economics} \cite{Fisher83}. This book
reviews a research program involving many prominent economists that attempted
to make models of the approach to equilibrium under the Walrasian framework.
They focused on the problem of \textit{tatonnement}, i.e. the process by which
prices arrive at their equilibrium values. This body of theory was not very
successful. In retrospect the difficulty was that the models required strong
and detailed assumptions about the market mechanism and the conclusions that
could be derived were not very general. This serves as a cautionary tale. The
hope for a new attempt is that we have new tools (the computer), much better
data, and perhaps a few new ideas.

Another often-used point of comparison between equilibrium in physics and
economics concerns power laws. As already mentioned, physical equilibrium does
not yield power laws. This can be explained fairly trivially: Under the Jaynes
formulation of statistical mechanics, deducing the correct probability
distribution of states is just a matter of maximizing entropy subject to any
imposed physical constraints, such as those that come from conservation laws.
So, for example, for a closed system with a fixed number of particles,
conservation of energy is just a constraint on the mean energy, and maximizing
entropy subject to this constraint yields an exponential distribution of
energies, which is not a power law. One can also work this out in more general
settings, e.g. when the energy or the number of particles is not conserved,
and a similar answer is obtained. It is not clear that this pertains in
economics. For example, doing the same calculation under a constraint on the
mean logarithm yields a power law; if there were economic systems where this
constraint were natural (e.g. due to the use of logarithmic utility), power
laws might emerge naturally. And as we have already stressed, in any case the
notions of equilibrium in economics and physics are quite different.

\subsection{The sub-prime mortgage crisis of 2007}

As we are writing this in the late summer of 2007, a financial crisis is
unfolding that provides a good illustration of the value of equilibrium
models, at the same time that it illustrates the need for non-equilibrium
models. The crisis began in sub-prime mortgages and appears to be spreading to
seemiingly unrelated sectors. This will take some time to understand
completely, but in cartoon form the crisis unfolded something like this:

\begin{itemize}
\item Low interest rates, a rising housing market, and confidence in the
risk-reducing power of mortgage-backed securities led to a proliferation of
sub-prime mortgages, often under questionable terms.

\item A drop in housing prices made it clear that many sub-prime borrowers
would eventually have to default.

\item This stressed capital providers and caused them to tighten credit
generally. One effect was that subprime borrowers themselves found it harder
to refinance, which paradoxically created more defaults and higher losses.

\item Another effect is that the prices of companies that depend on credit to
do business were depressed.

\item Institutions (including the largest banks in the United States) holding
AAA rated securities called CDOs backed by bonds which in turn were backed by
sub-prime mortgages lost something like \$100 or \$200 billion. Curiously
these banks announced their losses piecemeal. One week a bank would announce
\$10 billion of losses, the next week they would add another \$10 billion,
then two weeks later another \$10 billion.

\item In mid August 2007, in the midst of the sub-prime mortgage crisis, there
was a sudden meltdown of many statistical arbitrage hedge funds trading
equities. Stocks held by these hedge funds made unusual but systematic
relative movements -- stocks owned by these hedge funds dropped in value,
while those shorted by these funds rose in value\footnote{
This is based on anecdotal conversations with hedge fund managers. See also
\cite{Khandani07}.}.

\item One popular explanation is that the institutions losing money in
sub-prime mortgages sold positions in other markets, including stocks held by
stat arb hedge funds. This caused the stat arb hedge funds to take losses and
to liquidate more stocks, causing more losses and more liquidations.

\item The President announced that the crisis might be leading the country
into a recession.
\end{itemize}

This crisis involves some phenomena that are inherently out of equilibrium. At
the same time it shows the relevance of the equilibrium model.

The crisis was originally precipitated by levels of credit that are difficult
to justify as rational. Homeowners got several hundred thousand dollar loans
with no money down (i.e. the loan was equal to the appraisal value of the
house), and representations about income were taken at face value (i.e. the
income was not verified, but only stated by the homeowner himself). While
housing prices were rising, this caused no trouble. If the homeowner did not
pay, foreclosure could bring enough money to make the loan good, or more
common, its threat could induce the homeowner to refinance into another loan.
As housing prices started to decline, homeowners who in fact had no income
could not pay, and foreclosures are not expected to yield nearly enough cash
to repay the loans. 

As of this writing, the crisis is based entirely on expectations, including
expections about future home prices. The actual realized losses from
foreclosures so far have only been around 1\%. The \$250 billion loss that the
banks and hedge funds have been forced to take is based on the expectation
that eventually, in several years, 50\% of the homeowners (2.5 million
families) will be thrown out of their houses with losses of 50 cents on the
dollar for each foreclosed loan.\footnote{There is \$1 trillion in outstanding
subprime loans. Losses of $50\%\times50\%\times\$1~\mbox{trillion}=\$250$
billion.} The equilibrium model is thus right to put so much emphasis on the
importance of expectations. We have yet to see whether these expectations are rational.

One important contributing factor is that collateral levels have changed
dramatically. Before the crisis, one could get a sub-prime loan with almost no
money down. As the crisis developed, the required downpayment jumped to 25\%.
Since most sub-prime borrowers have little free cash, this meant that
effectively the sub-prime mortgage market dried up overnight. In the past,
subprime borrowers who made 36 straight payments, demonstrating they deserved
a better credit rating, would refinance into cheaper prime mortgages with
nothing down. The refinancing rate was over 70\% cumulatively for these
people. This is important because the subprime interest rates typically reset
to higher rates after 36 months. Now these same people are being asked to put
25\% down to refinance into a new mortgage. Many of these homeowners will not
be able to refinance, and they will eventually default, which will force
further foreclosures, further depressing housing prices.

The collateral tightening is spreading throughout the mortgage market.
Multibillion dollar hedge funds now have to put up twice as much cash to buy
mortgage derivatives as they had three months before (that is, instead of
borrowing 90\% of the price using a mortgage derivative as collateral, they
can only borrow 80\%).

A striking feature of the crisis is that the situation evolved rapidly and
appeared to be driven by emotion -- the word \textquotedblleft fear", which is
not an equilibrium concept, appeared in almost every newspaper article
covering these events. 

The crisis underlined the role of specialized players: banks, mortgage backed
derivatives experts, stock funds using valuation models, and quants building
statistical arbitrage strategies. Key aspects are the heterogeneous nature of
the strategies and their linkages to each other through institutional
groupings and cross-market exposure.

Is the crisis a simple consequence of irrationality, completely orthogonal to
the equilibrium model? It is not possible that lenders were so foolish that
they did not imagine the possibility that housing prices might fall. But they
apparently underestimated the amount of fraud in the loans. The brokers
arranging the loans did not own them. The loans were sold and repackaged into
bonds that were in turn sold to hedge funds and other investors. Since the
brokers knew they would not bear the losses from defaulted loans, they
evidently did not have enough incentive to check the truthfulness of the
borrowers. And the investors apparently did not monitor the brokers carefully
enough, perhaps lulled by the good behavior of the loans during the period of
housing appreciation. There was a breakdown of rationality.

The next incredible blunder was that the big banks bought so many
collateralized debt obligations (CDO's). To be sure, they only bought those
securities in the CDOs that were rated AAA by the rating agencies. But when
they do a deal the rating agencies are paid a commission for rating the
securities and if they do not approve enough AAA securities the deal does not
get done and they get no money. The rating agencies do not own the securities
themselves, and so they do not have the best incentive to get their ratings
right. It is amazing that the banks did not take this into account and naively
bought the AAA securities thinking they were truly safe.

Though many aspects of this crisis involve out-of-equilibrium phenomena like
irrational optimism, equilibrium theory can still be very insightful in
understanding what happened. In fact, it partially predicted the crisis.

In the first place the brokers who winked at bad loans while collecting
their\ commisions, and the rating agencies who gave high grades to bad bonds
while collecting their commisions, were both responding to their incentives,
just as equilibrium theory would predict, and as some economists had predicted.

But more importantly, the swift transition from lax collateral levels
permitting too many risky loans to overly tight collateral levels strangling
the sub-prime market is typical of the so-called leverage cycle in equilibrium
theory. Once the equilibrium model is extended to allow for default on
promises, and to allow for the posting of collateral to guarantee loans,
equilibrium determines the collateral levels that will be used for promises
(equivalently the leverage borrowers will choose). The kind of bad news that
increases uncertainty also increases equilibrium collateral levels. In normal
times these collateral levels are too lax, and in crisis times they are too
tight (see refs. \cite{Geanakoplos03,Fostel08}).

\section{If not equilibrium, then what?\label{beyond}}

Ken Arrow began a lecture he gave at the Santa Fe Institute a few years ago by
saying that ``economics is in chaos". In saying this he drew a contrast to the
situation in say, 1970, when rational expectations based equilibrium theories
were producing many new results and it looked as though they might be able to
solve many of the major problems in economics. At the time the path for a
young theoretical economist was clear. Now, in contrast, this is up in the
air. While the majority of economists still use rational expectations as the
foundation for what they do, even the mainstream is investigating
perturbations of this foundation, and some are seeking an entirely new
foundation. In this chapter we will sketch some possible directions. Many of
these approaches are overlapping and we try to stress the connections between them.

We are not suggesting that equilibrium theory should be thrown out: as
emphasized earlier, there are many situations where it is extremely useful.
What we do argue is that it should be one among many tools in the economist's
toolbox. During the rise of the neoclassical paradigm over the last fifty
years the emphasis on equilibrium theory has been so strong that most
theoretical economists have been trained in little else.\footnote{
The obvious exception is econometrics, which is an essential part of
economics, and should remain so. But econometric models are not founded on
\textit{a priori} models of agent behavior, and are not theories in the sense
that we are using the word here.}
As a result economics has suffered.

Disequilibrium models are not new in economics. Keynes, for example, made
models that were essentially dynamical, without imposing equilibrium
conditions. These models fell out of favor for the good reason that many of
them failed. In the 1950s and 1960s Keynesians accepted the inflation-output
trade-off we mentioned earlier, but government efforts to exploit that
relationship collapsed completely during the stagflation of the 1970s (when
there was both high inflation and low output). These failures were driving
forces in the modern equilibrium approach to macroeconomics, led by Lucas. It
is clear that for many problems in economics the ability to incorporate human
reasoning is necessary. However, there may be other ways to accomplish this
goal, and there may also be other important economic problems where human
reasoning is not the central issue.

\subsection{Behavioral and experimental finance}

As already stressed in Section~\ref{realism}, one of the principal problems
with rational expectations equilibrium is the lack of realism of the agent
model. The importance of work in this direction was acknowledged by the
mainstream with the Nobel prizes of Daniel Kahneman, who is one of the
pioneers behavioral economics, and Vernon Smith, who is one of the pioneers of
experimental economics. Because these important fields are already
well-covered in many review articles
\cite{Kagel97,Shefrin01,Barberis03,Thaler05} we will not review them here, but
rather only make a few comments.

We think that a proper characterization of human behavior is an essential part
of the future of economics. This is not going to be easy; people are
complicated and their behavior is malleable and context dependent.
Experimental economics offers hope for categorizing the spectrum of human
behavior and predicting how people will behave in given situations, but the
results are difficult to reduce to quantitative mathematical form. So far
behavioral finance has done a good job of documenting the many ways in which
real investors are not rational, and has shown that this has important
financial consequences. Although there has been some progress in understanding
the implications of these facts for classical problems such as saving and
asset pricing \cite{Berns07,Shefrin05}, the jury is still out concerning
whether this can be done in a fully quantitative manner.

One must also worry that behavioral finance might be the victim of its own
success. If we discover new empirical rules for the irrational component of
human behavior, might not these rules be violated as people become aware of
them? For example, it has been widely shown that people are strongly prone to
overconfidence. If this knowledge becomes widespread, will savvy investors
learn to compensate? In financial markets there is a lot of money on the table
and the motivation for overcoming irrational behavior is large. Of course, if
this were to occur that would already be a major achievement for the field.

\subsection{Structure vs. strategy}

In comparing theoretical economists to theoretical biologists, Paul Krugman
has commented that the two are not so different as it might seem: both make
extensive use of game theoretic models \cite{Krugman96}. This observation is
true, but it obscures a very important matter of degree. In biology game
theory models are one of many approaches used by theoreticians. In economics,
game theory models and their corollary form, equilibrium models, are almost
the only approach. From a certain point of view this is natural -- the need to
take into account strategic interactions in economics is larger than it is in
biology. The problem is that strategic interactions are not the only important
factor in economic models. Other factors can also be important, such as the
nature of economic institutions, and how the interactions of agents aggregate
to generate economic phenomena at higher levels. We will call the aspects of a
problem that do not depend on strategy its \textit{structure}.  In economics this occurs when equilibrium plays a role that is minor compared to other factors, such as interaction dynamics or budgetary constraints.  We give several examples below.  The substantial
effort needed to capture the strategic aspects of a problem in an equilibrium
model can often cause the structural aspects of a problem to be short-changed.
In some cases the importance of structural factors may dominate over strategic
considerations, and models that place too much emphasis on strategic
interactions without proper emphasis on structural properties will simply fail
to capture the essence of the problem. Of course in general one must take both
into account.

\subsubsection{Examples where structure dominates strategy}

As an example of what we mean, consider traffic. Cars are driven by people,
who anticipate the decisions of themselves and others. Yet an examination of
the literature on traffic and traffic control suggests that game theory and
equilibrium play only a minor role. Instead the dominant theories look more
like physics \cite{Helbing01,Nagel03}. Traffic is modeled by analogy with fluid flow.
The road is a structural constraint and cars are interacting particles.
Appropriately modified versions of the same ideas that are used to understand
the phases of matter in physics are used to understand traffic: A road with
only a few cars is like a gas, a crowded road where traffic is still flowing
is like a liquid, and a traffic jam is like a solid. In the liquid phase
perturbations can propagate and induce traffic jams; the equations that are
used to understand this are familiar to fluid dynamicists.\footnote{In the
medium density range, when drivers have delayed reactions or over-reactions to
random variations in the traffic flow, a breakdown in traffic flow is caused
by a dynamic instability of the stationary and homogeneous solution.} This is
not because equilibrium is not relevant, but rather because driving is tightly
constrained by roads, traffic control, driving habits, and response times. The
strategic considerations are fairly simple and their importance is dominated
by the mechanics of particles (automobiles) moving under structural constraints.

The same holds true for crowd dynamics, where game theoretic notions of
equilibrium would also seem to be important. In moving through a crowd one
tries to avoid collisions with one's neighbors, while trying to reach a given
goal, such as a subway entrance. One's neighbors are thinking about the same
thing. There are lots of strategic interactions and it would seem that game
theory should be the key modeling principle. In fact this does not seem to be
the most important consideration for modeling the movement of a crowd.

The problem of crowd dynamics becomes particularly important in extreme
situations when the density of people is high. For example, on January 12,
2006 roughly three million Muslims participated in the Hajj to perform the
stoning ritual. There was a panic and several hundred people died as a result
of either being crushed or trampled to death. In an effort to understand why
this occured the output of video cameras was processed in order to reduce the
crowd dynamics to a set of quantitative measurements. Analysis by Helbing,
Johannson and Al-Abideen \cite{Helbing07} showed that high densities of people
facilitate the formation of density waves, similar to those that occur in
driven granular media.\footnote{During a New Year's Eve celebration at
Trafalgar Square in London in 1980 one of the authors (jdf) personally
experienced the propagation of density waves through a large crowd. The
density waves propagate remarkably slowly. Their propagation speed can be
crudely estimated by regarding each person as an inverted pendulum.}
The crowd dynamics amplify small perturbations and create pressure fluctuations
that caused people to be crushed. The insights that were obtained by the study
of the Hajj data led to several modifications in the methods of crowd control,
and to a safe Hajj pilgrimage the following year.

The skeptical economist might respond that these are not economic problems,
and just because physical models such as fluid dynamics apply to crowd
behavior does not imply that they should be useful in economics. We included
the examples above because they provide clear examples of situations that
involve people, where despite the presence of strategic interactions, their
ability to reason is not a dominant point in understanding their aggregate
behavior. We now discuss a few problems in economics where similar
considerations apply.

\subsubsection{Distribution of wealth and firm size}

As originally observed by Pareto, the distribution of income displays robust
regularities that are persistent across different countries and through time
\cite{Pareto96}. For low to medium income it has a functional form that has
been variously described as exponential or log-normal
\cite{Dragulescu01,Yakovenko07}, but for very high incomes it is better
approximated by a power law. Since the early efforts of Champernowne, Simon,
and others, the most successful theories for explaining this have been random
process models for the acquisition and transfer of wealth
\cite{Champernowne53,Simon55,Malcai99,Sornette97,Dragulescu01,Levy05,}. If
these theories are right, then the distribution of wealth, which is one of the
most remarkable and persistent properties of the economy, has little to do
with the principles of equilibrium theory, and indeed little to do with human
cognition. Other problems with a similar flavor include the distribution of
the sizes of firms, the size of cities and the frequency with which words are
used \cite{Zipf32,Zipf49,Simon55,Ijiri77,Gabaix99,Axtell99,Axtell01}.

\subsubsection{Zero intelligence models of auctions}

The continuous double auction, which is perhaps the most commonly used price
formation mechanism in modern financial markets, provides a nice example of
how in some cases it can be very useful to focus attention on structural
rather than strategic properties. In a continuous double auction buyers and
sellers are allowed to place or cancel trading orders whenever they like, at
the prices of their choosing. If an order to buy crosses a pre-existing order
to sell, or vice versa, there is a transaction. The way in which orders are
placed is obviously closely related to properties of prices, such as
volatility and the spread (the price gap between the best standing sell order
and the best standing buy order). There have been many attempts to model the
continuous double auction using equilibrium theory. One famous example is the
Glosten model \cite{Glosten95}, which predicts the expected equilibrium
distribution of orders under a single period model. Tests of this model show
that it does not match reality very well \cite{Sandas01}.

In contrast, an alternative approach to this problem assumes a \textit{zero
intelligence} model.\footnote{
``Zero intelligence" loosely refers to models in which the agents have minimal cognitive capacity.  It should more accurately be properly called ``$\epsilon$ intelligence", since often agents do have some cognitive ability, e.g. the capacity to draw from a known distribution or to enforce a budget constraint.}
This approach was originally pioneered by Becker
\cite{Becker62}, who showed that some aspects of supply and demand curves
could be understood without any reliance on strategic thinking. The notion
that a zero intelligence model might be useful for understanding economic
phenomena was further developed by Gode and Sunder \cite{Gode93}, who
popularized this phrase and argued that observations of efficiency in
classroom experiments on auctions could be explained without relying on agent
intelligence. Zero intelligence models of the continuous double auction make
the assumption that people place orders more or less at random. We say ``more
or less" because there is more than one way to characterize their behavior,
and one can argue that some of these involve at least some strategic thinking,
and so are not strictly speaking zero intelligence. The development of zero
intelligence models of the continuous double auction has a long history
\cite{Mendelson82,Cohen85,Domowitz94,Bak97,Bollerslev97,Eliezer98,Maslov00,Slanina01,Challet01,Daniels03,Chiarella02,Bouchaud02,Smith03,Farmer05,Mike05}%
.

In some cases zero intelligence models make successful predictions. For
example, by assuming that order flow follows simple Poisson processes, the
model of Daniels et al. \cite{Daniels03,Smith03} derives equations of state
relating statistical properties of order flow such as the rates of trading
order placement and cancellation to statistical properties of prices such as
volatility and the bid-ask spread. Some of these predictions are borne out by
empirical data \cite{Farmer05}. The development of such zero intelligence
models has guided more empirically grounded \textquotedblleft low intelligence
models", which refine the statistical processes for order placement and
cancellation. For certain categories of stocks this results in accurate
quantitative prediction of the distribution of volatility and also of the
distribution of the spread \cite{Mike05}.

These models can be criticized because they fail in an important respect: the
resulting price series are not efficient. This is a failing, but it
illustrates an important point: the fact that a good prediction of the
distribution of volatility can be made from a model that does not satisfy
efficiency suggests that the distribution of volatility may not depend on
efficiency. Instead, the structural properties of the continuous double
auction appear to be more important. In this case by \textquotedblleft
structural properties" we mean those that come directly from the structure of
the auction, e.g. the rules under which transactions take place, the dynamics
of removal and deposition of orders, and their interactions with price
formation. These models suggest that for many purposes the structure of the
institution used to make trades plays an important role.

The zero intelligence approach is similar to the equilibrium model in the
sense that both approaches are parsimonious but sometimes highly unrealistic.
By making the simple assumption that people do not think at all, one can focus
on aspects of a problem that would be missed by an equilibrium analysis.
Equilibrium and zero intelligence models rovide ways to enter the wilderness
of bounded rationality without becoming totally lost; the two approaches are
complementary because they enter this wilderness from opposite sides.

Of course, many problems, including the double auction, require an
understanding of both structure and strategy, and the art of good modeling is
to find the right compromise in emphasizing these two facets of economics. A
good recent example is the theory of Wyart et al. who show that a mixture of
market efficiency and structural arguments can be used to explain the
relationship between the spread, price impact and volatility in transaction
time \cite{Wyart06}.

\subsection{Bounded rationality, specialization, and heterogeneous
agents\label{heterogeneousAgents}}

As originally pointed out by Adam Smith, the cognitive limitations of real
agents cause them to specialize.\footnote{
Equilibrium models often begin with asymmetric information a priori. See for
example the classic paper of Grossman and Stiglitz \cite{Grossman80}, which
can be viewed as an example of a rational treatment of heterogeneous agents
emerging from idiosyncratic information. Noise trader models provide another
example \cite{Schleifer00}. For a modern treatment of agents who are a priori
the same but choose to specialize and learn different things on account of
information processing constraints see \cite{Geanakoplos91}.}
In finance,
agents use specialized trading strategies. A few common examples are
fundamental valuation, technical trading (interpreting patterns in historical
price movements), many forms of derivative pricing, statistical arbitrage,
market making, index arbitrage, and term structure models. The champions of
heterogeneous models believe that to understand asset pricing as it manifests
itself in the real world, it is necessary to understand the behaviors of at
least some of these types of agents and their inter-relationships and interactions.

During the last couple of decades a new school of boundedly rational
heterogeneous agent (BRHA) models has emerged in finance. The primary
motivation of this work has been to explain phenomena such as bubbles and
crashes, clustered volatility, excess volatility, excess trading volume and
the heavy tails of price returns. As discussed in Sections 6
and~\ref{mayNotDepend}, these phenomena have so far not been explained by
equilibrium theory. In contrast there are now many BRHA models that produce
these phenomena. The typical BRHA models that have been constructed so far
study a single asset and assume that exogenous information enters through a
stochastic process corresponding to the dividends paid by that asset. The most
common technique is to simulate the resulting market behavior on a computer,
though some studies also derive analytic results. This style of work often
goes under the name of \textit{agent-based modeling}
\cite{LeBaron00,Tesfatsion99,tesfatsion05}.

Roughly speaking BRHA models can be divided into those that identify classes
of agents a priori and those that use learning to generate heterogeneous
agents de novo. Examples of models that assume heterogeneous agents a priori
are found in references
\cite{Beja80,Brock96,Brock97,Brock98,Caldarelli97,Chiarella92,Chiarella01,Day90,Farmer02,Farmer02b,Kirman91,Kirman93,Lux97,Lux98,Lux99,Rieck94}%
. The most commonly studied categories of agents are value investors, who
price stocks according to their fundamentals, and trend followers, who buy
stocks that have recently increased in price and sell stocks that have
recently decreased in price. The learning models assume heterogeneous learning
algorithms (or starting points) a priori and then let the strategies evolve as
learning takes place, thereby generating the heterogeneous strategies de novo.
Examples of this approach are given in references
\cite{Arthur97,Challet05,Giardina03,Johnson03,LeBaron99,LeBaron00,LeBaron01b,LeBaron03}%
. This approach lets strategies co-evolve in response to the conditions
created by each other, as well as in response to exogenously changing
fundamentals. The typical result produces a highly heterogeneous population of
trading strategies.

Simulations of such systems show that they naturally generate the apparently
nonequilibrium phenomena mentioned earlier, including clustered volatility,
heavy tails, bubbles and crashes, and excess trading. In most of these models
the primary cause of clustered volatility is changes in the populations of
different kinds of trading strategies. The population can shift due to
exogenous conditions, but most of the shifting appears to be largely random,
due to context-dependent fluctuations in the profitability of one group vs.
another. Changes in the exogenously provided fundamentals are amplified with
feedback. In some cases it is possible to show that in an analogous
equilibrium model none of these phenomena would arise
\cite{Arthur97,LeBaron99,LeBaron01b}. Furthermore, in some cases it is
possible to calibrate these BRHA models against real data and get a good fit
for properties such as the distribution of returns and the autocorrelation
function of volatility \cite{Johnson03,Giardina03,LeBaron03,Lux08,Lux08b}.
Thus the more developed models in this class make testable predictions. Note
that such studies do not claim that information arrival does not have an
effect, but only that a substantial component of volatility, and in particular
its interesting statistical properties, are internally generated.

Most of these BRHA models produce uncorrelated prices and arbitrage efficiency
between the simulated strategies. In the more sophisticated models, capital is
reallocated so that more profitable agents gain capital, giving them a greater
effect on price formation. After sufficient simulation time these models reach
a state in which the remaining strategies are all equally profitable (and are
in this sense arbitrage efficient). When this occurs the autocorrelations in
prices also disappear. This is not complete efficiency in the sense that it is
possible (at least in principle) to introduce new strategies that are not
within the generated set, that can exploit nonlinear price patterns and
temporarily make profits. (For some examples see \cite{Farmer02}.) After
enough time for capital to reallocate the advantage of those strategies also
disappears. This shows that that the prediction of efficiency by equilibrium
models is not as striking as it might seem -- the simulation models discussed
above generate results that are difficult to distinguish from true efficiency.
Thus, these models provide a means, albeit crude, to take reasoning into
account, and satisfy many of the desiderata for modeling agent behavior
discussed in Section~\ref{virtues}.

A great deal remains to be done in this area. Many of the the models mentioned
above illustrate the principle (by generating say clustered volatility), but
they do not make quantitative predictions about the real economy. Most of them
are not very parsimonious. There are still major open questions about the
necessary and sufficient conditions required for generating the phenomena of
interest. For instance, it is not absolutely clear what features of the models
that generate the phenomena are incompatible with equilibrium. It is even
possible that some equilibrium model might also generate the same phenomena,
but equilibrium models are so hard to compute that examples have not yet been
discovered. Further validation and testing and simplification are needed
before we can have a high degree of confidence that the detailed explanations
of these BRHA models are correct. Still more work is required to apply the
lessons of BRHA models to asset pricing and market design. Nonetheless, these
models apper to confirm the physics lesson that certain kinds of empirical
regularities (power laws) are the consequences of non-equilibrium behavior of
the market.

\subsection{Finance through the lens of biology}

The influence of biology on economics dates back at least to Alfred Marshall
\cite{Thomas91,Foster93,Nelson82,England94,Ruth96,Farmer99b,Farmer02,Lo05b}. Both biology and
economics involve specialized agents evolving through time under strong
selection pressure, and it is not surprising that they share many features,
such as extensive use of game theory. In this section we consider other
aspects of biology that are not captured by game theory and ask whether
financial economics might not benefit by taking some lessons from biology.

Note that the biological point of view is particularly complementary with that
of BRHA modeling. Biology is after all a study of specialized organisms and
their interactions with each other, and thus provides a lens for viewing many
aspects of financial markets.

\subsubsection{Taxonomy of financial strategies}

Taxonomy is the study of classification. By sorting complex phenomena such as
organisms into categories we compress an enormous amount of information into a
comprehensible scheme. Taxonomy has historically been one of the principal
activities of biologists. Perhaps the most famous taxonomist is Carl Linnaeus,
who in the middle of the eighteenth century introduced the hierarchical
classification of organisms into categories at varying levels of
differentiation, proposed a list of characteristics for classification, and
created the standardized nomenclature that is still largely used today. The
taxonomic classification of organisms has benefited from the participation of
hundreds of thousands or perhaps millions of individuals for more than three
hundred years and continues to be a field of active research and sometimes
heated debates in biology. Although taxonomy might seem to be a very limited
end in and of itself, it has proved to be an important prerequisite for the
development and testing of the theory of evolution in biology.

An economist, Adam Smith, was one of the first to point out the importance of
specialization, and in some branches of economics, such as industrial
economics, this is a field of active research. Not so in finance. To our
knowledge there are only two papers that gather empirical data on the nature
of real trading strategies in financial markets (and both of these papers only
address the frequency of usage of two preassigned categories of strategies,
technical trading and value investing) \cite{Keim95,Menkhoff98}. The diversity
and specialized functions performed by the trading desks of major financial
institutions suggest that there is an enormous amount waiting to be done if we
are to understand the taxonomy of real trading strategies. The BRHA models
mentioned above suggest that such diversity is important and plays a key role
in price formation. If this endeavor is to enter the realm of hard science it
needs a grounding in empirical data. We believe that market taxonomy should
become a major field of study by financial economists. Electronic data storage
presents the possibility of gathering data with information about the identity
of agents, which can potentially form the basis for such taxonomies.
Preliminary results in this direction have already yielded interesting results
\cite{Lillo07,Zovko07}.

\subsubsection{Market ecology}

For biological systems ecology is the study of organisms and their
relationship to their environment, including each other. Simulations of BRHA
models make it clear that understanding the relationships between financial
strategies is also important in financial economics. Market ecology is an
attempt to develop a theory that gives a deeper and more universal
understanding of this \cite{Farmer02}.

Under the analogy to biology a financial strategy can be viewed as a species,
and the capital invested in it can be viewed as its population. If strategies
are successful they accumulate more capital, which causes them to have more
influence in setting prices. The price provides the principal channel through
which different strategies interact -- the movements of prices influence the
buying and selling activity of strategies, which in turn affects prices. On a
longer timescale the movements of prices also affects profitability, which
alters the capital invested in strategies and is the main driver of the
selection process underlying market evolution.

In most BRHA models, price setting is done using a standard method, such as
market clearing, but even when this is the case, it can be very useful to view
the interactions of strategies in terms of their market impact. The market
impact is the price movement corresponding to the initiation of a trade of a
given size, and is equivalent (up to a constant of proportionality) to the
demand elasticity of price originally introduced by Marshall \cite{Farmer07b}.
Market impact is an increasing function of trading volume. As the capital
invested in a strategy grows, market impact lowers realized returns and
ultimately limits the trading capital that any given strategy can support.
With a proper understanding of market impact it is possible to derive
differential equations for the dynamics of capital flows between strategies,
analogous to the generalized Lotka-Volterra models of population biology
\cite{Farmer02}.

Market impact also provides a way to understand the strength of the
interactions between strategies and to classify them in ecological terms. In
particular, it is possible to make a pairwise classification of strategies
according to their influence on each other's profits by understanding how a
variation in the capital associated with strategy $j$ influences the profits
of strategy $i$. Letting $\rho_{i}$ be the return of strategy $i$ and $C_{j}$
be the capital invested in strategy $j$, one can define a gain matrix
$G_{ij}=\partial\rho_{i}/\partial C_{j}$. If $G_{ij}<0$ and $G_{ji}<0$
strategies $i$ and $j$ have a \textit{competitive} relationship; if $G_{ij}>0$
and $G_{ji}<0$ then they have a \textit{predator-prey} relationship, in which
$i$ preys on $j$, and if $G_{ij}>0$ and $G_{ji}>0$ they have a
\textit{mutualistic} (or symbiotic) relationship. The ecological view asserts
that the set of such relationships in the market is critical to its function,
and that the proper diversity of financial strategies may be critical for
market stability. This view also provides an evolutionary framework in which
to understand the path toward efficiency.

Until recently this ecological view of markets was essentially hypothetical
and it was not clear that the resulting models were testable. Recently,
however, the acquisition of data sets in which there is information about
brokerages or individual investors has opened up the possibility of
calibrating such models and testing their predictions against real data
\cite{Lillo07,Zovko07}. It will be interesting to determine whether one can
link properties of financial ecologies to properties of price formation. For
example, do alterations in financial ecologies influence price volatility?

\subsubsection{Evolutionary finance}

The three essential elements of evolutionary theory are descent, variation,
and selection. Financial ecologies are formed by these precisely these forces:

\begin{itemize}
\item \textit{Descent.} Financial strategies are passed down through time as
traders change and start new firms.

\item \textit{Variation.} Old strategies are modified and improved to create
new strategies.

\item \textit{Selection}. Successful strategies proliferate and unsuccessful
strategies disappear. Success is substantially based on profitability, but is
also influenced by other factors such as marketing and psychological appeal.
\end{itemize}

The view that profit selection is important is not new, having been championed
as a force for creating efficient markets by central figures in financial
economics such as Milton Friedman and Eugene Fama \cite{Friedman53, Fama70}.
The theory of market efficiency amounts to the assertion that it is possible
to short-circuit the evolutionary process and go directly to its endpoints, in
which evolution has run its course and profits of well-developed strategies
are reduced to low levels. The suggestion that the timescales for this process
are not short, as presented in Section~\ref{efficiencyProgression}, motivate
more detailed studies of the transient properties of the evolutionary process
and its dynamics. This view implicitly underlies all the BRHA models we have
discussed so far, and is also now being given a more mathematical underpinning
\cite{Hens05,Hens05b}.

What is still largely lacking is an empirical foundation on which to build an
evolutionary theory. This ultimately depends on developing a taxonomy of real
financial strategies together with a database of studies of financial
ecologies as they change through time.

\subsection{The complex systems viewpoint}

Complex systems refers to the idea that systems composed of simple components
interacting via simple rules can give rise to complex emergent behaviors, in
which in a certain sense the whole is greater than the sum of its parts. With
his introduction of the concept of the \textquotedblleft invisible hand" Adam
Smith was one of the earliest to articulate this point of view. The general
equilibrium theory of Arrow and Debreu can be regarded as an attempt to cut
through the complexity of individual interactions and reduce the invisible
hand to a tractable mathematical form. It appears, though, that for many
purposes this approach is just too simple -- to understand real financial
economies we need to do the hard work of understanding who the agents are,
what factors cause them to change, and how they interact with each other. In
order to do this we need better taxonomic studies of real markets, better
simulation models, and better theory. To the external observer, current
research in financial economics has so far largely failed to capture the
richness of real markets, which provide some of the best examples of complex
systems. More robust contact between financial economics and the emerging
field of complex systems could help remedy this situation.

\section{Conclusion\label{conclusions}}

We have written this article with a dual purpose. On one hand, we worry that
physicists often misunderstand the equilibrium framework in economics, and
fail to appreciate the very good reasons for its emergence. On the other hand,
the majority of economists have become so conditioned to explain everything in
terms of equilibrium that they do not appreciate that there are many
circumstances in which this is unlikely to be appropriate. We hope that
physicists will begin to incorporate equilibrium into their models when
appropriate, and that economists will become more aware of analogies from
other fields and begin to explore the possibilities of alternatives to the
standard equilibrium framework.

Our own belief is that one must choose modeling methods based on the context
of the problem. In situations where the cognitive task to be solved is
relatively simple, where there is good information available for model
formation, and where the estimation problems are tractable, rational
expectations equilibrium is likely to provide a good explanation. Some
examples where this is true include option pricing, hedging, or the pricing of
mortgage-backed securities. In other cases where the cognitive task is
extraordinarily complex, such as the pricing of a new firm, or where
estimation problems are severe, such as portfolio formation, human models may
diverge significantly from rational models, and the equilibrium framework may
be a poor approximation. For good science one must choose the right tool for
the job, and in this case the good scientist must use an assortment of
different tools. Close-mindedness in either direction is not likely to be productive.

As we have stressed, equilibrium theory is an elegant attempt to find a
parsimonious model of human behavior in economic settings. It can be
criticized, though, as a quick and dirty method, a heroic attempt to simplify
a complex problem. Now that we have begun to understand its limitations, we
must begin the hard work of laying new foundations that can potentially go
beyond it.

\section{References}

\bibliographystyle{acm}
\bibliography{jdf}

\section*{Acknowledgements}

\noindent We would like to thank Legg-Mason and Bill Miller for supporting the
conference ``Beyond Equilibrium and Efficiency", held at the Santa Fe
Institute in 2000, that originally stimulated this paper, and for supporting our subsequent research. This work was also supported by Barclays Bank and NSF grant HSD-0624351. Any opinions, findings
and conclusions or recommendations expressed in this material are those of the
authors and do not necessarily reflect the views of the National Science Foundation.

\end{document}